\documentclass[10pt, twocolumn, notitlepage, superscriptaddress, prb, longbibliography,floatfix]{revtex4-2}

\usepackage[utf8]{inputenc}
\usepackage[export]{adjustbox}

\usepackage{braket}
\usepackage{float}
\usepackage{amssymb}
\usepackage[intlimits]{amsmath}
\usepackage{graphicx}
\usepackage{here}
\usepackage{ulem}
\usepackage{tikz}
\usepackage{makecell}
\usepackage{array,varwidth,lipsum}
\usepackage{color}
\usepackage[colorlinks=true,linkcolor=Blue,citecolor=Blue,filecolor=Blue]{hyperref}

\hypersetup{
    colorlinks=true,
    urlcolor=blue,
    citecolor=blue,
    linkcolor=blue}

\begin{document}
\title{Engineering spin-orbit effects and Berry curvature by deposition of Eu on WSe$_{2}$}

\author{Johanna P. Carbone}
\email{j.carbone@fz-juelich.de}
\affiliation{Peter Gr\"unberg Institut and Institute for Advanced Simulation, Forschungszentrum J\"ulich and JARA, 52425 J\"ulich, Germany \looseness=-1}
\affiliation{Physics Department, RWTH-Aachen University, 52062 Aachen, Germany}

\author{Dongwook Go}
\affiliation{Peter Gr\"unberg Institut and Institute for Advanced Simulation, Forschungszentrum J\"ulich and JARA, 52425 J\"ulich, Germany \looseness=-1}
\affiliation{Institute of Physics, Johannes Gutenberg University Mainz, 55099 Mainz, Germany}

\author{Yuriy Mokrousov}
\affiliation{Peter Gr\"unberg Institut and Institute for Advanced Simulation, Forschungszentrum J\"ulich and JARA, 52425 J\"ulich, Germany \looseness=-1}
\affiliation{Institute of Physics, Johannes Gutenberg University Mainz, 55099 Mainz, Germany}

\author{Gustav Bihlmayer}
\affiliation{Peter Gr\"unberg Institut and Institute for Advanced Simulation, Forschungszentrum J\"ulich and JARA, 52425 J\"ulich, Germany \looseness=-1}

\author{Stefan Bl\"{u}gel}
\affiliation{Peter Gr\"unberg Institut and Institute for Advanced Simulation, Forschungszentrum J\"ulich and JARA, 52425 J\"ulich, Germany \looseness=-1}

\begin{abstract}
Motivated by recent progress in 2D spintronics, we present Eu deposited on a 1H-WSe$_2$ as a promising platform for engineering spin-orbit effects and Berry curvature. By first-principles calculations based on density functional theory, we show that Eu/WSe$_2$ exhibits intriguing properties such as high magnetic anisotropy, valley-dependent polarization of spin and orbital angular momenta, and their Rashba textures. These originate from magnetic and spin-orbit proximity effects at the interface and the interplay between localized $4f$ magnetic moments of Eu and mobile charge carriers of WSe$_2$. We find a pronounced anomalous Hall effect in the proposed system. Thus, we promote $4f$ rare-earth metals deposited on top of a transition-metal dichalcogenides as a promising platform for 2D spintronics.
\end{abstract}
\maketitle
\section{Introduction}
Transition-metal dichalcogenides (TMDCs) are 2D-materials with formula MX$_{2}$, where M is a transition-metal element bound to chalcogen atoms X. Based on the composition ($i.e.$ the nature of M and X) and the crystalline structure \cite{Lee2018, Chhowalla2015} (the stacking sequence in the bulk systems), such systems exhibit different electronic properties and can be semi-metals, semiconductors, metals, or even superconductors \cite{PRB2011,TMDC2006,PhysRevLett2010, 2011PhyB, NbSe2_super,TMDCphyschem,naturecomm, naturecomm2018,NbSeTe}. Moreover, individual monolayers of TMDC are atomically thin structures, which can manifest either the hexagonal symmetry (1H-phase) or the octahedral symmetry (1T-phase) \cite{Chhowalla2015}. These materials exhibit interesting phenomena such as valley degrees of freedom \cite{liu2019valleytronics} and band splitting \cite{zibouche2014transition} due to the presence of different features they naturally possess, like inversion symmetry breaking and strong spin-orbit coupling (SOC). The inversion symmetry breaking leads to valley-dependent orbital splittings at the corner points $\mathrm{K}$ and $\mathrm{K}'$ of the hexagonal Brillouin zone (BZ), which are characterized by the opposite signs of the orbital angular momentum and Berry curvature. This leads to valley-orbital Hall effect, where electrons at $\mathrm{K}$ and $\mathrm{K}'$ valleys carry the opposite orbital angular momentum \cite{mak2014valley,PhysRevLett.99.236809, Bhowal2020, Canonico2020, Cysne2021, Vignale2021}. Furthermore, strong SOC leads to valley-dependent spin polarization and results in the spin Hall effect \cite{PhysRevLett.108.196802}. The valley-dependent Berry curvature also leads to the Berry curvature dipole, which drives a nonlinear Hall effect \cite{Nature2019,Liang2015,Xie2018,Qin_2021}. 

In recent years, different studies on TMDCs have been carried out, showing how the symmetry properties have an important impact on the transport properties. In Ref. \cite{PRB2020} it is demonstrated that by application of an electric field perpendicular to a WSe$_{2}$ bilayer, structural inversion symmetry is broken and it is thus possible to control the Berry curvature and the orbital moment at the $\mathrm{K}$ and $\mathrm{K}'$ valleys by tuning the external electric field. Reducing the system to a monolayer, different kinds of SOC effects arise due to the reduction in symmetry which can be exploited in valleytronics as explained in Ref. \cite{CommPhys2019}. 
By proximity effect it is also possible to induce exchange interaction by depositing a ferromagnetic layer on top of a TMDC. Since a clean surface of TMDCs can be prepared by exfoliating, it can intimately contact with a ferromagnet \cite{zhao2017enhanced}. It is found that these magnetic interactions modify the extent of the Hall effects compared to the situation without exchange interaction  \cite{JAP2020}. 
The coexistence of strong SOC, emerging from the transition-metal of a TMDC monolayer, and of a net magnetization is also the key ingredient for magnetotransport phenomena such as the anomalous Hall effect (AHE) \cite{PRB2017} and spin-orbit torque \cite{Hidding2020,Gambardella2011}. Especially, low symmetry of the TMDCs can induce unconventional spin polarization and torque \cite{MacNeill2017, Hidding2020, Xue2020}, which is crucial to achieve field-free switching of the magnetization.

By similar reasoning, the presence of rare-earth adatoms with high coverage on top of TDMCs is here demonstrated to produce an additional contribution to the Hall conductivity that depends on the particular topology of the band structure and can be described in terms of Berry curvature.
The combination of rare-earth atoms with 2D materials is a promising strategy for the implementation of novel magnetic storage devices and for applications in the field of spintronics. Indeed, the usage of rare-earths as a magnetic source has different advantages such as high magnetic moments generated by localized $4f$ electrons and strong SOC, whose effect together with the particular crystal field of the 2D material generates magnetocrystalline anisotropy, which is being intensively studied from both theoretical and experimental point of view \cite{Nature2013, Science2016, Nano2016, Shick2018, Shick2019}. This effect can be exploited for example in the creation of nanoscale magnets where the main challenge is the stabilization of the magnetic moments which undergo fluctuations due to vibrations or interaction with conduction electrons of the substrate. This can be achieved in several ways, for instance through manipulation of the symmetry properties of the system.


In the present work, the effect of the adsorption of Eu atoms on a WSe$_{2}$ monolayer in the 1H-phase is investigated through $ab$ $initio$ density functional theory (DFT) calculations. We calculate the magnetic ground-state simulating different spin-spirals and determine the magnetic anisotropy energy. A further analysis concerns the impact of the $4f$-metal on the spin-orbital texture and on the consequences of the interaction between the rare-earth electrons and the substrate on the band topology.
In particular, anomalous Hall transport properties are analyzed by considering a high coverage of the $4f$-metal and using interpolation by Wannier interpolation. It is shown that the adsorption of Eu generates states in the gap of the semiconducting substrate which are influenced by the SOC arising from the W atom and that hybridization between the different species induces a non-trivial Berry curvature and an anomalous Hall conductivity, which is experimentally measurable.

\section{Results}

\subsection{Computational Details}

\begin{figure}[b]
    \centering
    \includegraphics[scale=0.25]{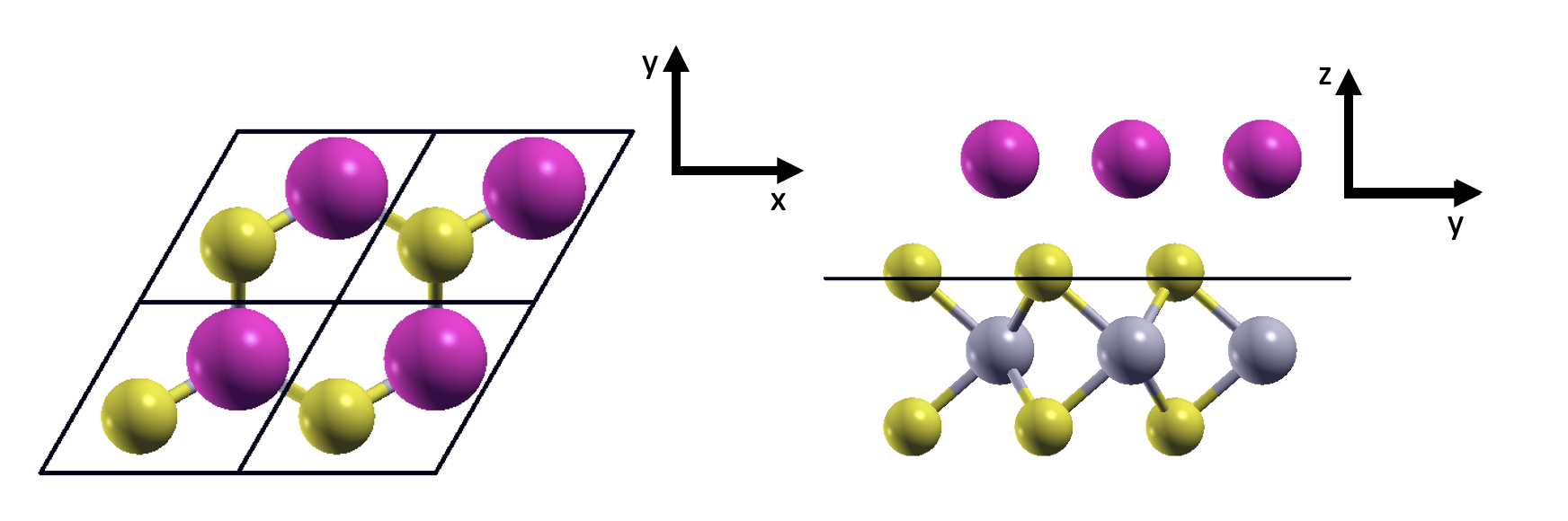}
    \caption{Structure of the $1\times 1$ unit cell of Eu (purple spheres) deposited on top of the W atom (grey spheres) of WSe$_{2}$. Se atoms are indicated by yellow spheres.}
    \label{unitcell}
\end{figure}

WSe$_{2}$ has a hexagonal structure characterized by one W atom covalently bonded together to six Se atoms. To determine the adsorption site for the Eu adatom, we compare the total energies of the three different sites: on top of the W atom (T-W), on-top of the Se atom (T-Se) and in the middle of the hexagon formed by W and Se atoms (H). The relaxation procedure and the following calculations are performed inside of a $1\times 1$ simulation cell with lattice constant $a=3.327$ \r{A} using the full-potential linearized augmented planewave (FLAPW) method as implemented in the FLEUR code \cite{FLEUR} and using DFT plus Hubbard U method \cite{DFTUShick} in order to account for the highly localized $4f$ electrons. In this respect, the Perdew, Burke, and Ernzerhof exchange-correlation functional \cite{PBE} is adopted and the onsite Coulomb and Hund exchange parameters are set to $U=6.7$ eV and $J=0.7$ eV, which are values generally accepted for half-filled $4f$-shells \cite{DFTUShick,Kurz_2002}. Concerning the DFT parameters, we set muffin-tin radii $2.80 a_0$ for Eu and $2.29 a_0$ for W and Se, where $a_0$ is the Bohr radius. The cut-off for the plane-wave basis functions is chosen to be $K_\mathrm{max}=4.0 a_0^{-1}$ and $G_\mathrm{max}=10.7 a_0^{-1}$. The upper limit of the angular momentum inside of the muffin-tin is set to $l_\mathrm{max}=10$ for Eu and  $l_\mathrm{max}=8$ for W and Se. Furthermore, a $10\times 10$ $\mathbf{k}$-point mesh is sampled throughout the first Brillouin zone for the self-consistent field cycle. 

Table \ref{table1} summarizes the adsorption energy of the Eu adatom on each adsorption site, the distance along the $z$ direction $i.e.$ perpendicularly from the top Se layer, the magnetic moment of Eu and the $f$ and $d$ occupations in the valence shell of Eu. The reference point of the energy is defined as the total energy for an isolated Eu monolayer and WSe$_2$. In the case of T-Se, $h$ represents the distance of the adatom from the Se which is directly underneath the Eu. From these data it is evident that the most stable adsorption site is on top of the W atom (Fig.\ref{unitcell}) and that the $4f$-metal maintains its high magnetic moment around $7.1$ $\mu_{B}$. The small deviation from the theoretical value suggested by Hund's rule, arises from a gain in the $5d$ occupation by proximity.

\begin{table}
\begin{tabular}{|c|c|c|c|c|c|}
\hline
$Site$ & $E_\mathrm{ads}$ (eV)  & $h$ (\r{A}) & $m_s^\mathrm{tot}$ ($\mu_{B}$)  & $f_\mathrm{occ}$ & $d_\mathrm{occ}$\\ \hline
H & $-0.312$ & $2.830$ & $7.240$ &$6.861$ & $0.520$\\ \cline{1-6} 
T-W & $-0.474$  & $2.500$ &  $7.130$ & $6.865$ & $0.522$ \\  \cline{1-6} 
T-Se & $-0.341$ & $3.119$  &  $7.440$ & $6.858$& $0.550$\\ \hline
\end{tabular}
\caption{Adsorption energy, distance of the adatom from the WSe$_{2}$ monolayer, the magnetic moment and the $f$ and $d$ occupations of the Eu adatom for the different adsorption sites in the $1\times 1$ unit cell.}
\label{table1}
\end{table}
The DFT energy bands are compared to the band structure obtained by constructing maximally-localized Wannier functions (MLWFs) in the FLAPW formalism \cite{Freimuth_2008} and the open-source code Wannier90 \cite{Wannier90}. The initial projections for the Wannier functions are chosen to be $s,d,f$ orbitals for the Eu atom, $p$ orbitals for the Se atoms and $s,d$ orbitals for the W atom. In this way, $50$ MLWFs are constructed, where the frozen window maximum was set 0.4 eV above the Fermi energy. From the converged MLWFs, the Hamiltonian, spin, and orbital operators are written in real space, which is Fourier-transformed in an interpolated $\mathbf{k}$-mesh for the calculation of spin-orbital texture, Berry curvature, and AHE.

\subsection{Electronic Structure}


\begin{figure*}[t]
    \centering
    \includegraphics[scale=0.4]{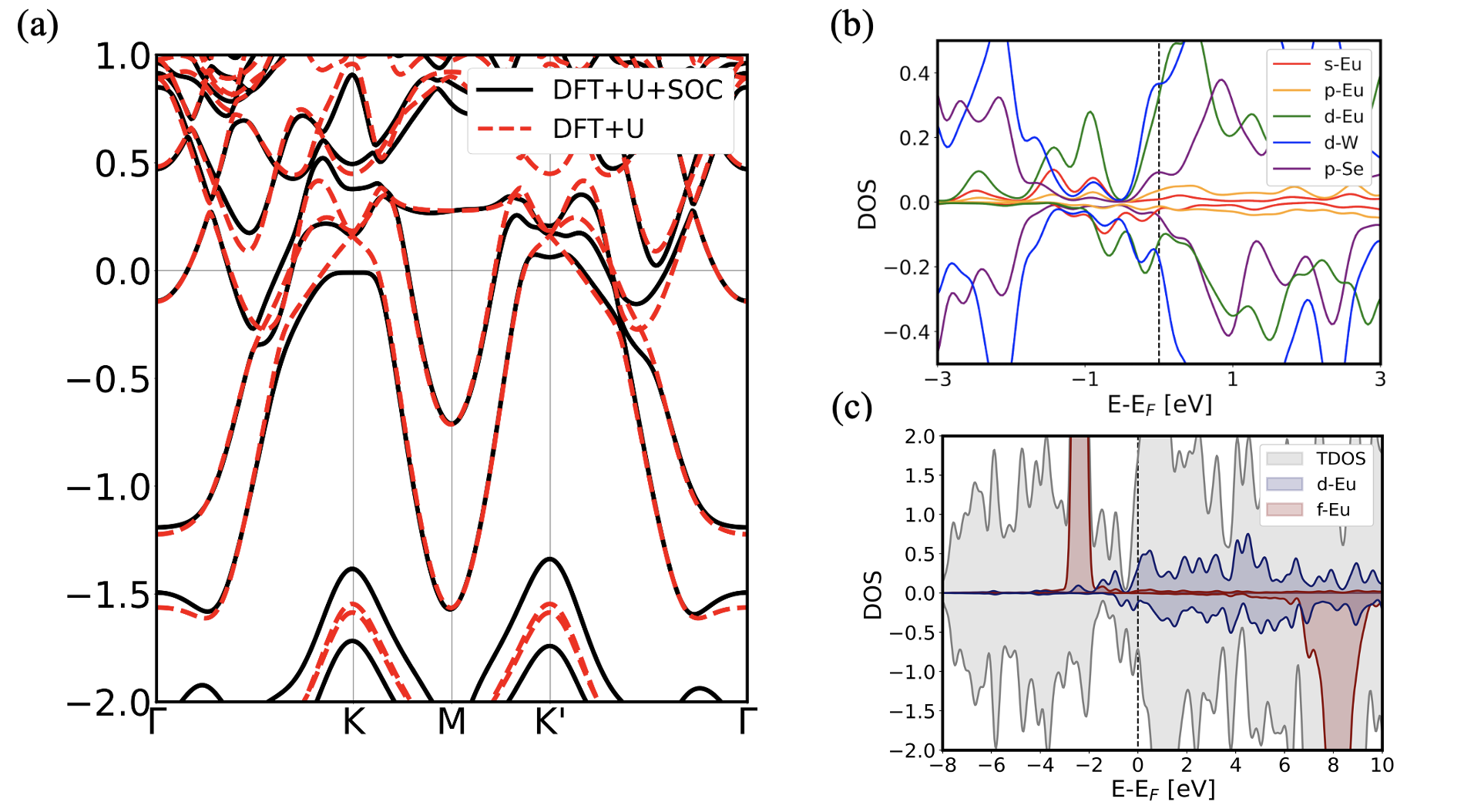}
    \caption{(a) Band structure of Eu/WSe$_{2}$ calculated with DFT+U (red dashed line) and with DFT+U+SOC (black solid line). (b) Contribution to the DOS of the $s,p,d$ electrons of Eu, $d$ electrons of W and $p$ electrons of Se. (c) Contribution to the DOS of the $f$ and $d$ electrons of Eu. The total DOS (TODS) is shown as grey shaded area.}
    \label{electronicstructure}
\end{figure*}


The effect of the interactions between adatom and substrate can be noticed by looking at the electronic structure of the material, shown in Fig. \ref{electronicstructure}. One immediately notices that the system is metallic while an isolated WSe$_2$ is insulating. This is due to hybridization between WSe$_2$ substrate and Eu atoms by proximity. Figure \ref{electronicstructure}(a) displays a comparison of the band structure with and without SOC. The SOC effect arises mainly from the W atom and it is seen how this affects states around the Fermi energy which from Fig.~\ref{electronicstructure}(b) can be understood to emerge from mixing of the energy levels of the different species. In particular, the valley-shaped states at $-2$ eV at the $\mathrm{K}$ point, which have predominantly $d$ character from W as seen in blue in Fig.~\ref{electronicstructure}(b), split in presence of SOC. In the same way also the crossing point at $\mathrm{K}$ just above the Fermi energy is split up into two separated bands. Another important feature arises on the path between $\mathrm{K}$ and $\Gamma$ on the Fermi energy, where SOC generates an avoided crossing. This plays a crucial role in generating large Berry curvature and contributing to the AHE. Concerning the energy bands from $\mathrm{K}'$ to $\Gamma$, similar effects arise but it can be already noticed that the two high symmetry points $\mathrm{K}$ and $\mathrm{K}'$ are not equivalent due to the absence of inversion symmetry.
Figure~\ref{electronicstructure}(b) shows the calculated contribution to the density of states (DOS) of $s,p$ and $d$ electrons of Eu, the $d$ electrons of W and the $p$ electrons of Se around the Fermi energy. It is seen that in this energy window these electrons with different atomic and orbital characters hybridize while the $4f$ electrons of Eu are localized at around $-3$ eV and $+8$ eV [Fig.~\ref{electronicstructure}(c)], which are responsible for the induced magnetization in the system. Because of the atomic-like nature of such electrons, the hybridization effects with the environment are in general small and limited to indirect interactions through itinerant electrons like the $d$ states of Eu.

\begin{figure}[b!]
    \centering
    \includegraphics[scale=0.42]{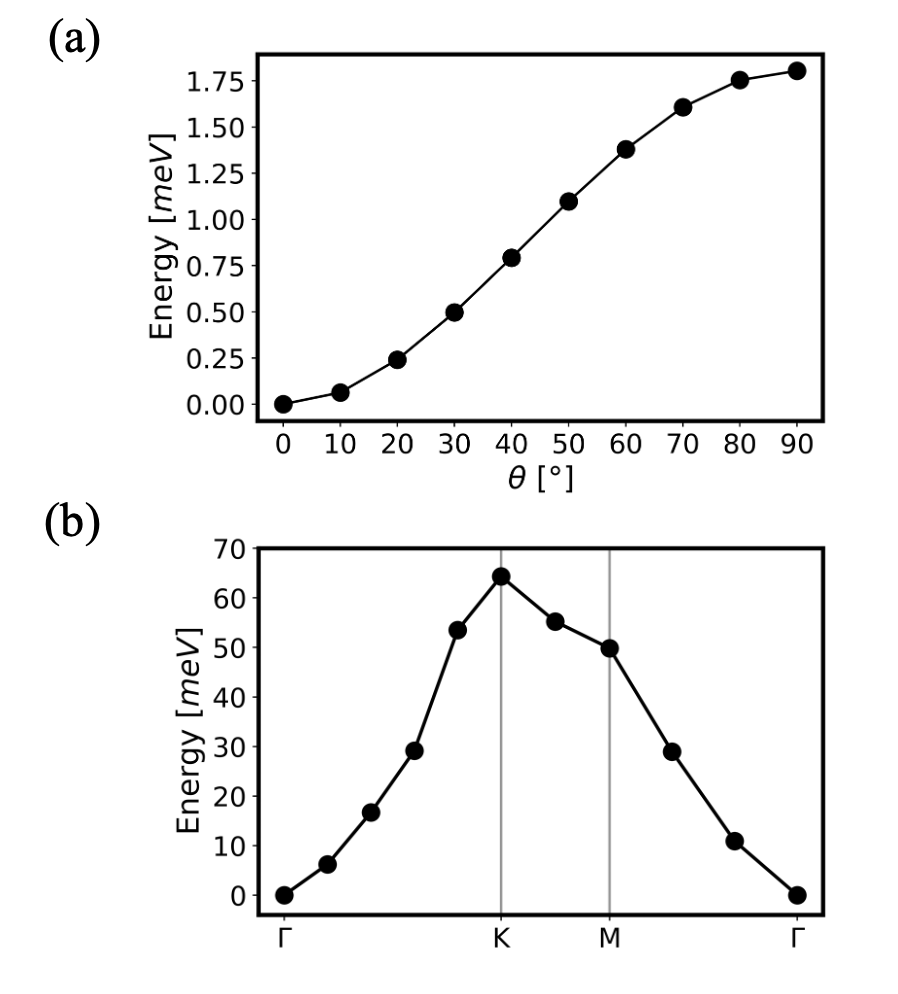}
    \caption{(a) Magnetic anisotropy energy curve: the total energy of the system plotted versus the angle $\theta$ of the magnetization with the $z$-axis. (b) The energy of the spin-spiral states with cone angle $\beta=\pi/2$ computed for the values of the $q$-vector along the  $\Gamma$-$\mathrm{K}$-$\mathrm{M}$ path, presented with respect to the ferromagnetic ground state at the $\Gamma$-point. }
    \label{spinspiral}
\end{figure}

\subsection{Perpendicular Magnetic Anisotropy}

\begin{figure*}[t!]
    \centering
    \includegraphics[scale=0.32]{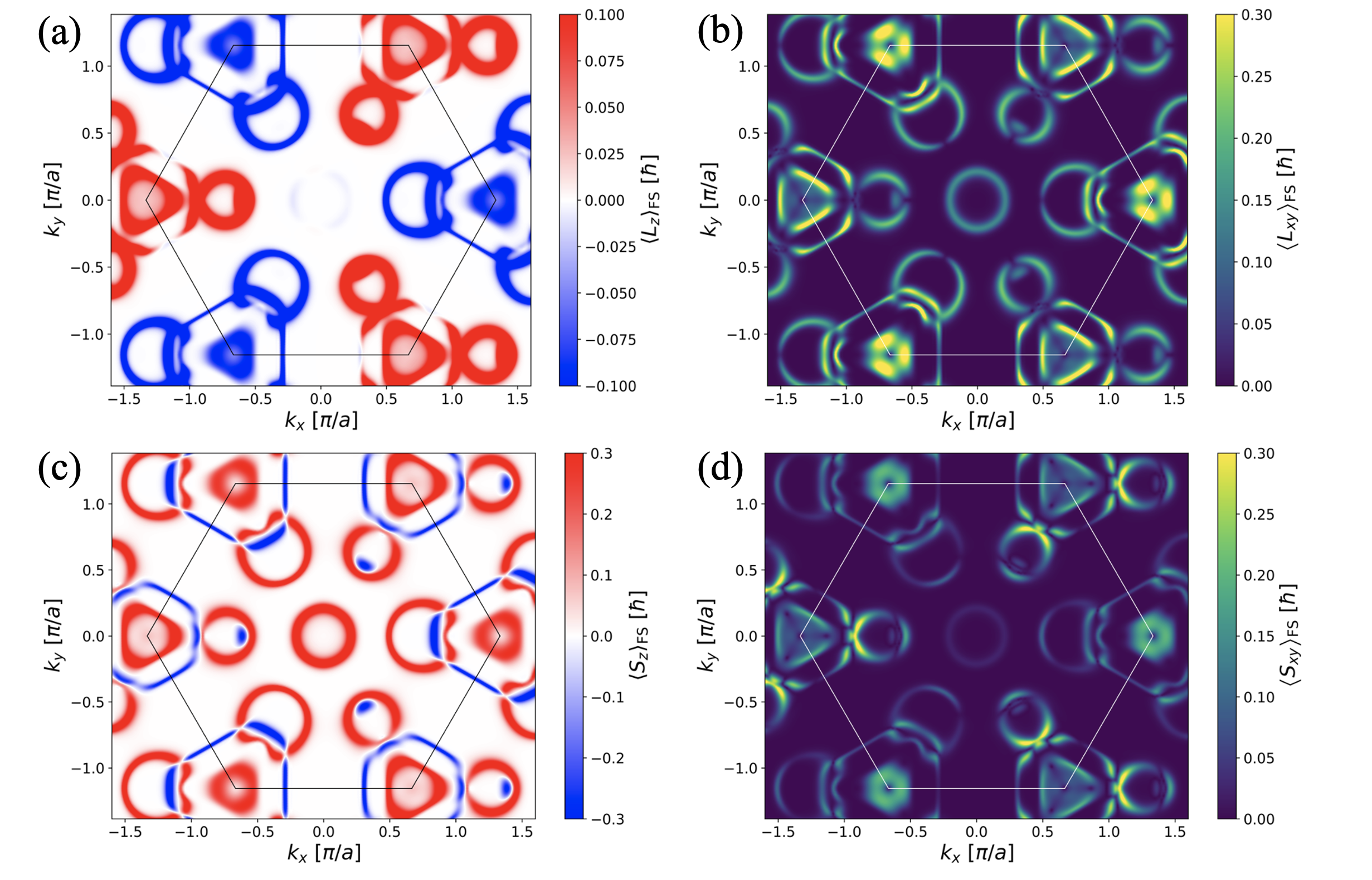}
    \caption{Spin and orbital texture in $\mathbf{k}$-space at the Fermi surface. (a) Expectation value for the out-of-plane component of the orbital angular momentum at the Fermi surface $\langle L_z \rangle_\mathrm{FS}$, (b) Magnitude of the expectation value of the in-plane component of the orbital angular momentum $\langle L_{xy} \rangle_\mathrm{FS} = \sqrt{\langle L_x \rangle_\mathrm{FS}^2 + \langle L_y \rangle_\mathrm{FS}^2}$. Analogously, the $z$-component and the magnitude of the in-plane component for the spin expectation value at the Fermi surface are shown in (c) and (d), respectively.}
    \label{Texture}
\end{figure*}

In magnetic data-storage devices, an important key ingredient is the stiffness of the magnetization with respect to external perturbations ($i.e$ thermal fluctuations or scattering from conduction electrons) such that a specific direction of the magnetic moments is stable over time, which translates into the need of high magnetic anisotropy energies. This is particularly important in thin film magnets, where high anisotropy can circumvent the Mermin-Wagner theorem in 2D \cite{Mermin1966, Griffiths1964}. In Fig.~\ref{spinspiral}(a), the total energy of Eu/WSe$_2$ is plotted with respect to the angle $\theta$ between the surface normal direction and the magnetization. 
The result indicates an out-of-plane easy axis of the magnetization with an energy difference of $1.75$ meV per unit cell compared to the energy of the in-plane state. 

This effect can be described in terms of magnetic anisotropy emerging from the charge transfer between Eu and the WSe$_2$. Eu has $7$ $4f$-electrons leading to a closed-shell situation for which the total orbital angular momentum is $L=0$ but the total spin moment is close to $7$ $\mu_{B}$ as discussed before. The hybridization with the  electrons of the substrate induces an anisotropic spin density, which deviates from the perfect spherical symmetry of an isolated atom. Together with the SOC, the crystal field couples to the spin density. Thus, the magnetic anisotropy emerges as a result of proximity-induced hybridizations.


To confirm whether the magnetic ground state is indeed a perpendicular ferromagnetic phase, we investigate the first-order energy correction from the SOC for a given spin spiral with a wave vector $\mathbf{q}$ (defined in terms of reciprocal lattice vectors) \cite{Kurz2004, Heide2009} by employing the generalized Bloch theorem \cite{rado1966magnetism, Sandratskii1986}. Figure~\ref{spinspiral}(b) shows the energy calculated for different values of the $\mathbf{q}$-vector of a spin spiral along the $\Gamma$-$\mathrm{K}$-$\mathrm{M}$ path. We find the energy minimum at $\Gamma$, indicating that the system favors a ferromagnetic ground state. Furthermore, the energy differences between the $\Gamma$ and the other two high-symmetry points, $\mathrm{K}$ (N\'eel state) and $\mathrm{M}$ (antiferromagnetic state) are about $70$ meV and $100$ meV. Thus, a N\'eel or antiferromagnetic state is not energetically favorable.


\subsection{Orbital and Spin Textures}

In a TMDC monolayer, spatial inversion symmetry is broken and the direct consequence is emergence of inequivalent valleys $\mathrm{K}$ and $\mathrm{K}'$ in $\mathbf{k}$-space. This leads to valley-dependent orbital angular momentum and Berry curvature and results in the valley-orbital Hall effect \cite{Bhowal2020, Vignale2021, Cysne2021}. Additionally, by depositing Eu atoms on WSe$_2$, the Rashba effect can be induced by breaking the mirror symmetry with respect to the 2D plane \cite{Rashba1984, Manchon2015}. The Rashba effect is induced not only on the spin but also on the orbital angular momentum, which is known as the orbital Rashba effect \cite{Park2011, Park2013, Go2017, Sunko2017}. These spin and orbital textures play a crucial role in magneto-transport phenomena of spin and orbital \cite{Go2018, Canonico2020, Go2021}.

Thus, we investigate the orbital and spin texture in $\mathbf{k}$-space at the Fermi surface of Eu/WSe$_2$. This is evaluated by
\begin{eqnarray}
\label{eq:orbital_texture}
\mathbf{L}_\mathrm{FS}(\mathbf{k})
&=&
\sum_n \frac{2
\bra{u_{n\mathbf{k}}} \mathbf{L} \ket{u_{n\mathbf{k}}}
}{1+\cosh [(E_\mathrm{F}-E_{n\mathbf{k}})/k_\mathrm{B}T]}
,
\end{eqnarray}
where $u_{n\mathbf{k}}$ is a periodic part of the Bloch state with band index $n$, $E_{n\mathbf{k}}$ is the corresponding energy band, $\mathbf{L}$ is the orbital angular momentum operator defined within muffin spheres of each atoms, and $E_\mathrm{F}$ is the Fermi energy. We set $k_\mathrm{B}T=25\ \mathrm{meV}$ for broadening, where $k_\mathrm{B}$ is the Boltzmann constant and $T$ is the temperature. Figure~\ref{Texture}(a) shows the $z$ component of the orbital angular momentum at the Fermi surface. We find a $3$-fold rotational symmetry, as expected, and the valley-dependent orbital texture is observed, which have opposite signs at $\mathrm{K}$ and $\mathrm{K}'$ valleys. We find that this feature is similar to bare WSe$_2$. However, by Eu adsorption in-plane components of the orbital angular momentum emerge, which are chiral in $\mathbf{k}$-space. In Fig.~\ref{Texture}(b), the magnitude of the in-plane components of the orbital angular momentum is shown, which is defined by $\langle L_{xy} \rangle_\mathrm{FS} = \sqrt{\langle L_x \rangle_\mathrm{FS}^2 + \langle L_y \rangle_\mathrm{FS}^2}$. We find a clear in-plane orbital texture which satisfies a three-fold rotation symmetry. We also find that its magnitude is different at each valley $\mathrm{K}$ and $\mathrm{K}'$. 

Analogously, we also calculate the spin texture by replacing $\mathbf{L}$ by $\mathbf{S}$ in Eq.~\eqref{eq:orbital_texture}. The out-of-plane and in-plane components of the spin texture are shown in Figs.~\ref{Texture}(c) and (d), respectively. Although they satisfy a three-fold rotation symmetry, as expected, the out-of-plane component shown in Fig.~\ref{Texture}(c) exhibit the distribution that is close to a hexagonal symmetry, where impact of the broken inversion symmetry is not evident. This is because the spin magnetism is mainly driven by Eu layer, which has a six-fold rotation symmetry if the substrate is absent. Slight deviation of the spin texture from the six-fold rotation symmetry indicates hybridization of Eu atoms with the substrate, where proximity-induced states exhibit finite spin polarization via an indirect exchange interaction between itinerant electrons and spin moments of localized $f$ electrons. In contrast to the out-of-plane component, the in-plane component shown in Fig.~\ref{Texture}(d) shows clear difference at $\mathrm{K}$ and $\mathrm{K}'$ valleys. This is because its origin is the orbital texture induced by hybridization between the substrate and Eu atoms [Fig.~\ref{Texture}(b)], which is manifested by the SOC that entangles the orbital and spin wave functions.




\subsection{Berry Curvature}
The features of the spin-orbital texture and the perpendicular magnetization arranged in a ferromagnetic fashion are an interesting starting point for the investigation of the anomalous conductivity. The latter can be described in terms of Berry curvature, which acts in $\mathbf{k}$-space as an effective magnetic field and causes a transverse electron current in presence of an electric field. The Berry curvature is evaluated using the Kubo formula,
\begin{eqnarray}
\Omega_{n\mathbf{k}}
&=&
\partial_{k_x} A_{n\mathbf{k}}^y - \partial_{k_y} A_{n\mathbf{k}}^x
\nonumber
\\
&=&
-2\mathrm{Im}
\left[
\braket{
\partial_{k_x} u_{n\mathbf{k}} |
\partial_{k_y} u_{n\mathbf{k}}
}
\right]
\nonumber
\\
&=&-2\hbar^2 \sum_{m\neq n}
\textrm{Im}
\left[
\frac{
\bra{u_{n\mathbf{k}}} v_x \ket{u_{m\mathbf{k}}}
\bra{u_{m\mathbf{k}}} v_y \ket{u_{n\mathbf{k}}}
}{(E_{n\mathbf{k}}-E_{m\mathbf{k}}+i\eta)^2}
\right],
\nonumber
\\
\label{Berrycurveq}
\end{eqnarray}
where $\Omega_{n\mathbf{k}}$ is the Berry curvature for a Bloch state with band index $n$, $\mathbf{A}_{n\mathbf{k}}=i\braket{u_{n\mathbf{k}} | \partial_\mathbf{k} u_{n\mathbf{k}}}$ is the Berry connection, and $v_{x(y)}$ is the $x(y)$ component of the velocity operator. We introduce a small positive number $\eta$, which is set to 25 meV for convergence. From Eq.(\ref{Berrycurveq}) it is clear that there will be contributions to the Berry curvature from the regions where the energy bands are separated by small energy gaps by the effect of SOC, such as avoided crossings discussed in Fig.~\ref{electronicstructure}(a). 

\begin{figure}[t!]
    \includegraphics[scale=0.35]{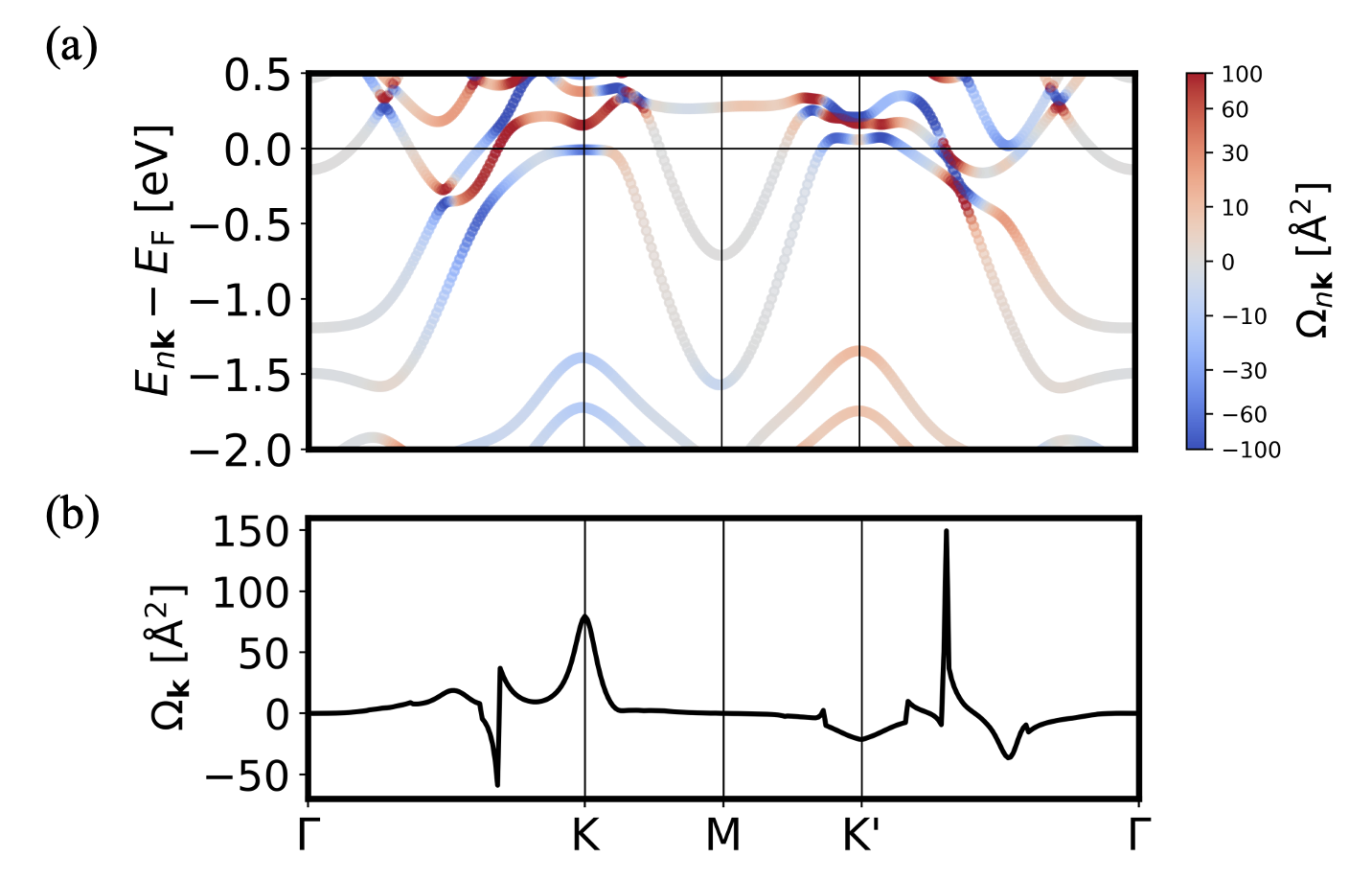}
    \caption{(a) Band structure around the Fermi energy with color scale indicating the value of the Berry curvature $\Omega_{n\mathbf{k}}$. (b) Berry curvature summed over all occupied states along the $\mathbf{k}$-path $\Gamma-\mathrm{K}-\mathrm{M}-\mathrm{K}'-\Gamma$.}
    \label{berry}
\end{figure}

While the Berry curvature vanishes when the spatial inversion and time-reversal symmetries are combined, in Eu/WSe$_2$ both symmetries are broken. In particular, orbital hybridizations by proximity and the SOC can generate strong Berry curvature near avoided crossings of bands as shown in Fig.~\ref{electronicstructure}(a). This explains the features of the Berry curvature shown in Fig. \ref{berry}(a), where the calculated band structure along the $\mathbf{k}$-path $\Gamma-\mathrm{K}-\mathrm{M}-\mathrm{K}'-\Gamma$ and the respective value is shown in color scale.
Different hotspots of $\Omega_{n\mathbf{k}}$ can be seen at points where the SOC lifts the degeneracy of the energy bands. In particular, the splitting of the bands between $\Gamma$ and $\mathrm{K}$ appears to result in a band inversion. In terms of Eq.~\eqref{Berrycurveq}, these splittings correspond to small values of the denominator and thus sharp peaks of the Berry curvature. In Fig.~\ref{berry}(b), we show the Berry curvature summed over all occupied states below the Fermi energy for the same $\mathbf{k}$-path. We confirm that spiky contributions comes from the SOC-induced avoided crossings, which is found on $\Gamma-\mathrm{K}$ and $\Gamma-\mathrm{K}'$ paths. Another important feature is that also in terms of Berry curvature, the two $\mathrm{K}$ and $\mathrm{K}'$ points are inequivalent: the $\mathrm{K}$ point presents a positive peak, while at the $\mathrm{K}'$ point displays a broad negative feature. Along the path $\mathrm{K}'-\Gamma$ an intense peak appears characterized by inverted sign with respect to the peak between $\Gamma-\mathrm{K}$. 

\subsection{Anomalous Hall Effect}

\begin{figure}[t!]
\centering
    \includegraphics[scale=0.42]{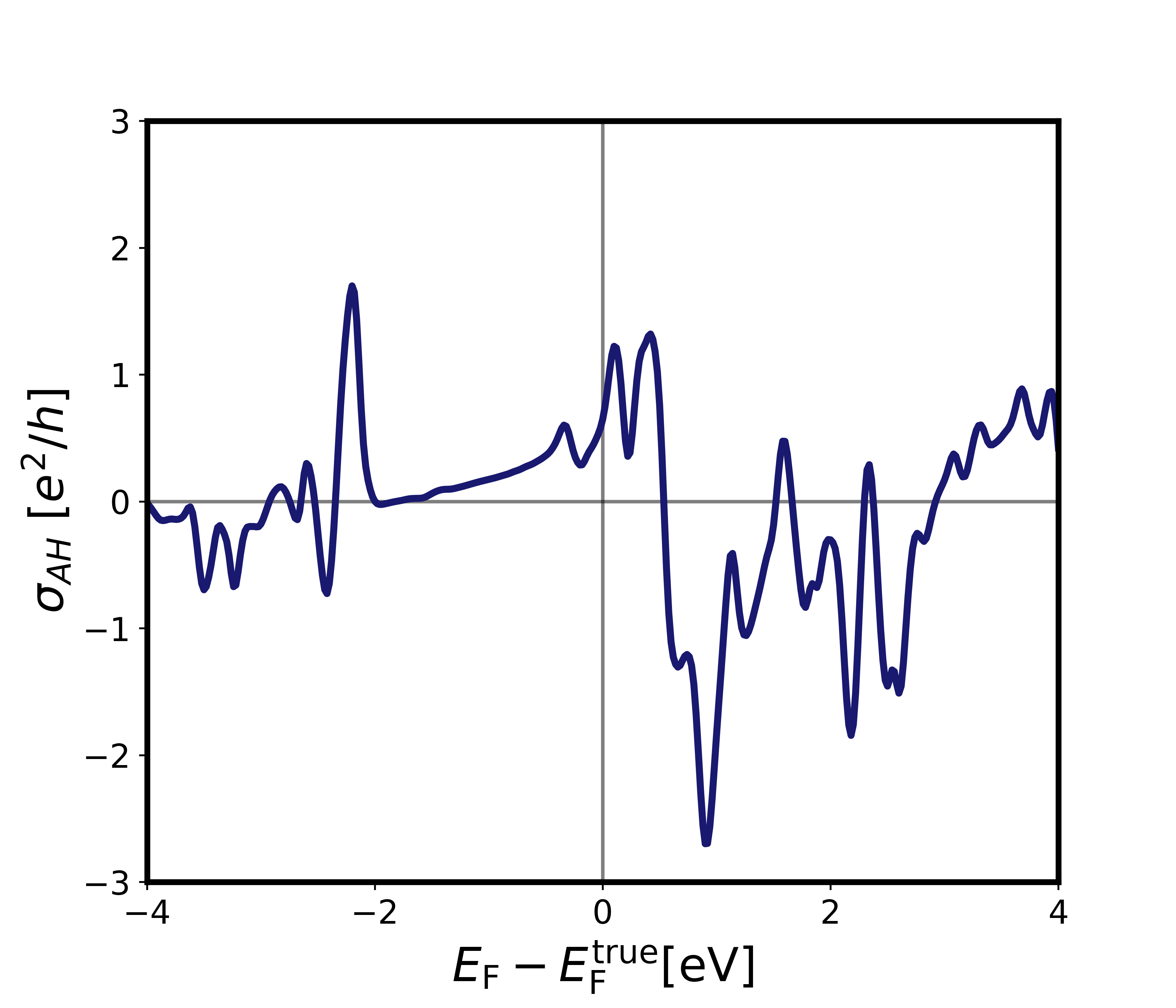}
    \caption{Anomalous Hall conductivity as a function of the Fermi level.}
    \label{AHC}
\end{figure}

By integrating the Berry curvature over the Brillouin zone (BZ), it is possible to calculate the intrinsic anomalous Hall conductivity as
\begin{equation}
    \sigma_\mathrm{AH}\equiv \sigma_{yx} = 
    \frac{e^{2}}{\hbar}\sum_{n}\int_{\textrm{BZ}} \frac{d^2k}{(2\pi)^{2}}f_{n\mathbf{k}} {\Omega}_{n\mathbf{k}},
\end{equation}
\noindent
where $f_{n\mathbf{k}}$ is the Fermi-Dirac distribution function. Figure~\ref{AHC} shows the anomalous conductivity $\sigma_{yx}$ as a function of the Fermi energy. The Fermi energy is varied with respect to the original value $E_\mathrm{F}^\mathrm{true}$ by assuming that the potential is fixed for the change of band filling. Major peaks are found around 2 eV below the Fermi energy, slightly above the Fermi energy, and 1 eV above the Fermi energy. These energies are where avoided band crossings induced by the SOC are found. A double-peak feature right above the Fermi energy implies an interesting possibility to tune the Hall response by electron doping, which may be experimentally observed. Meanwhile, the peak at $-2$ eV is where the $\mathrm{K}$ and $\mathrm{K}'$ valleys of WSe$_{2}$ in $\mathbf{k}$-space are situated. From Fig.\ref{electronicstructure}(a) it is clear that SOC lifts the degeneracy of the band at the  $\mathrm{K}$ and $\mathrm{K}'$ valleys which have predominantly $d$ character from the W atom such that a contribution to the Berry curvature arises. 

\section{Discussion}

\subsection{Effect of Eu Coverage}

\begin{figure}[t!]
    \includegraphics[scale=0.36]{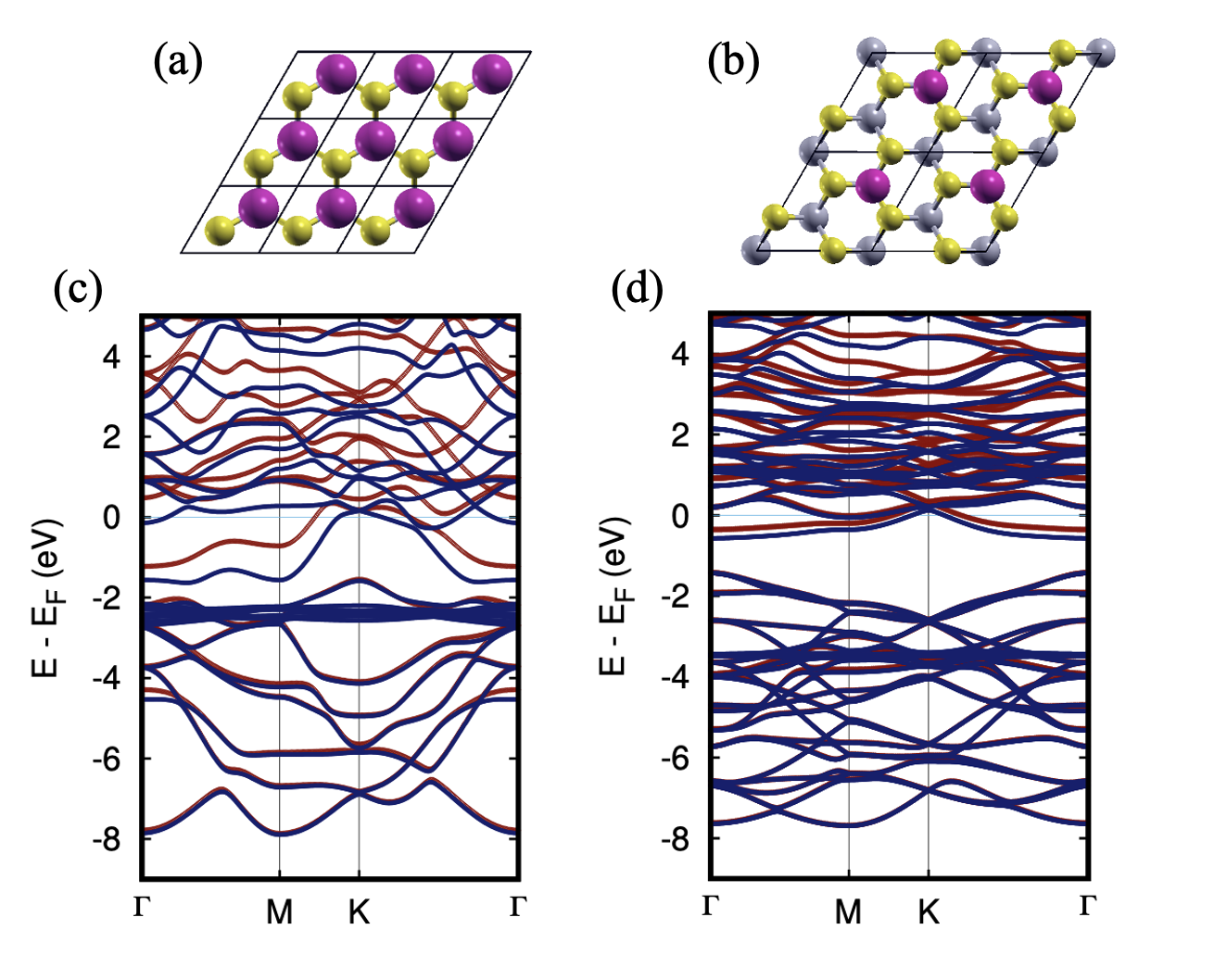}
    \caption{Comparison of the two simulation cells: (a) high coverage of Eu in $1\times 1$ unit cell and (b) low coverage of Eu in $\sqrt{3}\times\sqrt{3}$ unit cell. The corresponding band structures are shown in (c) and (d), where red and blue lines indicate majority and minority states, respectively.
    }
    \label{comparison}
\end{figure}

In the experimental setup, Eu atoms might not fully cover the WSe$_2$ substrate. To investigate an effect of Eu coverage, we also compare the electric structures of $1\times 1$ unit cell and $\sqrt{3}\times\sqrt{3}$ unit cell. As shown in Fig.~\ref{comparison}(a) and (b), the Eu coverage of the $1\times 1$ unit cell is three times higher than that of the $\sqrt{3}\times\sqrt{3}$ unit cell. In the above, we presented the result on $1\times1$ unit cell (1 magnetic atom per unit cell). This choice was driven by the observation that in a bigger simulation cell (lower coverage of the adatom on the substrate) the density of states of Eu around the Fermi energy is reduced. This can be seen from the band structures shown in Figs.~\ref{comparison}(c) and (d) for the $1\times1$ cell and the $\sqrt{3}\times\sqrt{3}$ cell, respectively. The SOC is not included in the calculation of these band structures. While we find proximity magnetism in both cases, locally for W atoms right below Eu atoms, the $\sqrt{3}\times\sqrt{3}$ case still presents an energy gap. This is in contrast to the $1\times 1$ case, which is metallic. As discussed in Fig.~\ref{electronicstructure}(a), metallicity is driven by the orbital hybridization between Eu atoms and WSe$_2$ substrate. This has a strong implication for a realization in experiments and suggests that high coverage of Eu atoms is crucial to be able to measure the AHE. This also implies that the AHE is expected to be enhanced as the coverage of Eu atoms increases.

\subsection{Other Rare-Earth Elements}

In order to tailor an efficient device, it is necessary to protect the perpendicular magnetization from perturbations that might flip it to in-plane. An enhancement of the anisotropy energy might be obtained for example by substituting Eu with an open $4f$-shell rare-earth (such as Nd, Dy or Ho) in which the charge cloud presents deviations from the spherical geometry and electron correlations play an important role.  In these cases the $4f$-shell gives rise to an anisotropic charge cloud which depends on the non-vanishing orbital angular momentum $L$. The big values of $L$ and $S$ are source of SOC and cause magnetocrystalline anisotropy which can be exploited if put in specific chemical environments. The task is to engineer properly the combination rare-earth/2D material in order to achieve high energy differences between different magnetic states. A theoretical challenge is to achieve an accurate description of open-shell $4f$ systems for which  approaches like DFT+U, self-interaction corrected DFT or dynamical mean field theory are adopted. A further approach is the Hubbard-I Approximation that has been implemented in recent times in several codes and applied to a variety of rare-earth systems \cite{Shick2018, Shick2019,ShickHoPt(111), Locht2018}. These approaches serve as future topics of investigation. 

\section{Conclusion}

In conclusion, we analyzed the effect of depositing Eu adatoms on top of a WSe$_{2}$ monolayer and showed how the interplay of proximity-induced orbital hybridization, SOC, magnetism, and broken symmetry leads to the possibility of engineering the spin-orbital texture and the Berry curvature in such heterostructures. We predict that hybridization effects between the electrons of Eu and the electrons of WSe$_{2}$ induce a magnetic anisotropy of about $1.75$ meV to rotate the magnetization from out-of-plane to in-plane. In addition, the analysis of the magnetic texture of the material for various spin spiral structures predicts that a ferromagnetic configuration is favoured. As a consequence of the synergy between these magnetic features and the particular spin-orbital texture, Berry curvature hotspots are induced in the $\mathbf{k}$-space, which in turn lead to anomalous Hall conductivity in the sample. Such rare-earth based technology can be exploited for spintronic devices with 2D materials.

\begin{acknowledgements}
We acknowledge Gregor Michalicek for giving advice on first-principles calculation. The project is funded by the DFG funded CRC 1238 – Control and Dynamics of Quantum Materials: Spin orbit coupling, correlations, and topology (Project C01). D.G. and Y.M. acknowledge funding by the Deutsche Forschungsgemeinschaft (DFG, German Research
Foundation) - TRR 288 -  422213477 (project B06). We acknowledge computing resources granted by RWTH Aachen University under project jara0219 and by the J\"ulich Supercomputing Centre under project jiff40. 
\end{acknowledgements}

\bibliography{main.bib}

\begin{thebibliography}{60}%
\makeatletter
\providecommand \@ifxundefined [1]{%
 \@ifx{#1\undefined}
}%
\providecommand \@ifnum [1]{%
 \ifnum #1\expandafter \@firstoftwo
 \else \expandafter \@secondoftwo
 \fi
}%
\providecommand \@ifx [1]{%
 \ifx #1\expandafter \@firstoftwo
 \else \expandafter \@secondoftwo
 \fi
}%
\providecommand \natexlab [1]{#1}%
\providecommand \enquote  [1]{``#1''}%
\providecommand \bibnamefont  [1]{#1}%
\providecommand \bibfnamefont [1]{#1}%
\providecommand \citenamefont [1]{#1}%
\providecommand \href@noop [0]{\@secondoftwo}%
\providecommand \href [0]{\begingroup \@sanitize@url \@href}%
\providecommand \@href[1]{\@@startlink{#1}\@@href}%
\providecommand \@@href[1]{\endgroup#1\@@endlink}%
\providecommand \@sanitize@url [0]{\catcode `\\12\catcode `\$12\catcode
  `\&12\catcode `\#12\catcode `\^12\catcode `\_12\catcode `\%12\relax}%
\providecommand \@@startlink[1]{}%
\providecommand \@@endlink[0]{}%
\providecommand \url  [0]{\begingroup\@sanitize@url \@url }%
\providecommand \@url [1]{\endgroup\@href {#1}{\urlprefix }}%
\providecommand \urlprefix  [0]{URL }%
\providecommand \Eprint [0]{\href }%
\providecommand \doibase [0]{https://doi.org/}%
\providecommand \selectlanguage [0]{\@gobble}%
\providecommand \bibinfo  [0]{\@secondoftwo}%
\providecommand \bibfield  [0]{\@secondoftwo}%
\providecommand \translation [1]{[#1]}%
\providecommand \BibitemOpen [0]{}%
\providecommand \bibitemStop [0]{}%
\providecommand \bibitemNoStop [0]{.\EOS\space}%
\providecommand \EOS [0]{\spacefactor3000\relax}%
\providecommand \BibitemShut  [1]{\csname bibitem#1\endcsname}%
\let\auto@bib@innerbib\@empty
\bibitem [{\citenamefont {Kim}\ and\ \citenamefont {Lee}(2018)}]{Lee2018}%
  \BibitemOpen
  \bibfield  {author} {\bibinfo {author} {\bibfnamefont {J.}~\bibnamefont
  {Kim}}\ and\ \bibinfo {author} {\bibfnamefont {Z.}~\bibnamefont {Lee}},\
  }\href {https://doi.org/10.9729/AM.2018.48.2.43} {\bibfield  {journal}
  {\bibinfo  {journal} {Applied Microscopy}\ }\textbf {\bibinfo {volume}
  {48}},\ \bibinfo {pages} {43} (\bibinfo {year} {2018})}\BibitemShut {NoStop}%
\bibitem [{\citenamefont {Voiry}\ \emph {et~al.}(2015)\citenamefont {Voiry},
  \citenamefont {Mohite},\ and\ \citenamefont {Chhowalla}}]{Chhowalla2015}%
  \BibitemOpen
  \bibfield  {author} {\bibinfo {author} {\bibfnamefont {D.}~\bibnamefont
  {Voiry}}, \bibinfo {author} {\bibfnamefont {A.}~\bibnamefont {Mohite}},\ and\
  \bibinfo {author} {\bibfnamefont {M.}~\bibnamefont {Chhowalla}},\ }\href
  {https://doi.org/10.1039/C5CS00151J} {\bibfield  {journal} {\bibinfo
  {journal} {Chem. Soc. Rev.}\ }\textbf {\bibinfo {volume} {44}},\ \bibinfo
  {pages} {2702} (\bibinfo {year} {2015})}\BibitemShut {NoStop}%
\bibitem [{\citenamefont {Kuc}\ \emph {et~al.}(2011)\citenamefont {Kuc},
  \citenamefont {Zibouche},\ and\ \citenamefont {Heine}}]{PRB2011}%
  \BibitemOpen
  \bibfield  {author} {\bibinfo {author} {\bibfnamefont {A.}~\bibnamefont
  {Kuc}}, \bibinfo {author} {\bibfnamefont {N.}~\bibnamefont {Zibouche}},\ and\
  \bibinfo {author} {\bibfnamefont {T.}~\bibnamefont {Heine}},\ }\href
  {https://doi.org/10.1103/PhysRevB.83.245213} {\bibfield  {journal} {\bibinfo
  {journal} {Phys. Rev. B}\ }\textbf {\bibinfo {volume} {83}},\ \bibinfo
  {pages} {245213} (\bibinfo {year} {2011})}\BibitemShut {NoStop}%
\bibitem [{\citenamefont {Wilson}\ and\ \citenamefont
  {Yoffe}(1969)}]{TMDC2006}%
  \BibitemOpen
  \bibfield  {author} {\bibinfo {author} {\bibfnamefont {J.}~\bibnamefont
  {Wilson}}\ and\ \bibinfo {author} {\bibfnamefont {A.}~\bibnamefont {Yoffe}},\
  }\href {https://doi.org/10.1080/00018736900101307} {\bibfield  {journal}
  {\bibinfo  {journal} {Advances in Physics}\ }\textbf {\bibinfo {volume}
  {18}},\ \bibinfo {pages} {193} (\bibinfo {year} {1969})}\BibitemShut
  {NoStop}%
\bibitem [{\citenamefont {Mak}\ \emph {et~al.}(2010)\citenamefont {Mak},
  \citenamefont {Lee}, \citenamefont {Hone}, \citenamefont {Shan},\ and\
  \citenamefont {Heinz}}]{PhysRevLett2010}%
  \BibitemOpen
  \bibfield  {author} {\bibinfo {author} {\bibfnamefont {K.~F.}\ \bibnamefont
  {Mak}}, \bibinfo {author} {\bibfnamefont {C.}~\bibnamefont {Lee}}, \bibinfo
  {author} {\bibfnamefont {J.}~\bibnamefont {Hone}}, \bibinfo {author}
  {\bibfnamefont {J.}~\bibnamefont {Shan}},\ and\ \bibinfo {author}
  {\bibfnamefont {T.~F.}\ \bibnamefont {Heinz}},\ }\href
  {https://doi.org/10.1103/PhysRevLett.105.136805} {\bibfield  {journal}
  {\bibinfo  {journal} {Phys. Rev. Lett.}\ }\textbf {\bibinfo {volume} {105}},\
  \bibinfo {pages} {136805} (\bibinfo {year} {2010})}\BibitemShut {NoStop}%
\bibitem [{\citenamefont {{Ding}}\ \emph {et~al.}(2011)\citenamefont {{Ding}},
  \citenamefont {{Wang}}, \citenamefont {{Ni}}, \citenamefont {{Shi}},
  \citenamefont {{Shi}},\ and\ \citenamefont {{Tang}}}]{2011PhyB}%
  \BibitemOpen
  \bibfield  {author} {\bibinfo {author} {\bibfnamefont {Y.}~\bibnamefont
  {{Ding}}}, \bibinfo {author} {\bibfnamefont {Y.}~\bibnamefont {{Wang}}},
  \bibinfo {author} {\bibfnamefont {J.}~\bibnamefont {{Ni}}}, \bibinfo {author}
  {\bibfnamefont {L.}~\bibnamefont {{Shi}}}, \bibinfo {author} {\bibfnamefont
  {S.}~\bibnamefont {{Shi}}},\ and\ \bibinfo {author} {\bibfnamefont
  {W.}~\bibnamefont {{Tang}}},\ }\href
  {https://doi.org/10.1016/j.physb.2011.03.044} {\bibfield  {journal} {\bibinfo
   {journal} {Physica B Condensed Matter}\ }\textbf {\bibinfo {volume} {406}},\
  \bibinfo {pages} {2254} (\bibinfo {year} {2011})}\BibitemShut {NoStop}%
\bibitem [{\citenamefont {Staley}\ \emph {et~al.}(2009)\citenamefont {Staley},
  \citenamefont {Wu}, \citenamefont {Eklund}, \citenamefont {Liu},
  \citenamefont {Li},\ and\ \citenamefont {Xu}}]{NbSe2_super}%
  \BibitemOpen
  \bibfield  {author} {\bibinfo {author} {\bibfnamefont {N.~E.}\ \bibnamefont
  {Staley}}, \bibinfo {author} {\bibfnamefont {J.}~\bibnamefont {Wu}}, \bibinfo
  {author} {\bibfnamefont {P.}~\bibnamefont {Eklund}}, \bibinfo {author}
  {\bibfnamefont {Y.}~\bibnamefont {Liu}}, \bibinfo {author} {\bibfnamefont
  {L.}~\bibnamefont {Li}},\ and\ \bibinfo {author} {\bibfnamefont
  {Z.}~\bibnamefont {Xu}},\ }\href {https://doi.org/10.1103/PhysRevB.80.184505}
  {\bibfield  {journal} {\bibinfo  {journal} {Phys. Rev. B}\ }\textbf {\bibinfo
  {volume} {80}},\ \bibinfo {pages} {184505} (\bibinfo {year}
  {2009})}\BibitemShut {NoStop}%
\bibitem [{\citenamefont {Ataca}\ \emph {et~al.}(2012)\citenamefont {Ataca},
  \citenamefont {Şahin}, ,\ and\ \citenamefont {Ciraci}}]{TMDCphyschem}%
  \BibitemOpen
  \bibfield  {author} {\bibinfo {author} {\bibfnamefont {C.}~\bibnamefont
  {Ataca}}, \bibinfo {author} {\bibfnamefont {H.}~\bibnamefont {Şahin}}, ,\
  and\ \bibinfo {author} {\bibfnamefont {S.}~\bibnamefont {Ciraci}},\ }\href
  {https://doi.org/10.1021/jp212558p} {\bibfield  {journal} {\bibinfo
  {journal} {The Journal of Physical Chemistry C}\ }\textbf {\bibinfo {volume}
  {116}},\ \bibinfo {pages} {8983} (\bibinfo {year} {2012})}\BibitemShut
  {NoStop}%
\bibitem [{\citenamefont {Li}\ \emph {et~al.}(2016)\citenamefont {Li},
  \citenamefont {Duerloo}, \citenamefont {Wauson},\ and\ \citenamefont
  {Reeda}}]{naturecomm}%
  \BibitemOpen
  \bibfield  {author} {\bibinfo {author} {\bibfnamefont {Y.}~\bibnamefont
  {Li}}, \bibinfo {author} {\bibfnamefont {K.-A.~N.}\ \bibnamefont {Duerloo}},
  \bibinfo {author} {\bibfnamefont {K.}~\bibnamefont {Wauson}},\ and\ \bibinfo
  {author} {\bibfnamefont {E.~J.}\ \bibnamefont {Reeda}},\ }\href
  {https://doi.org/10.1038/ncomms10671} {\bibfield  {journal} {\bibinfo
  {journal} {Nat. Commun.}\ }\textbf {\bibinfo {volume} {7}},\ \bibinfo {pages}
  {10671} (\bibinfo {year} {2016})}\BibitemShut {NoStop}%
\bibitem [{\citenamefont {Ciarrocchi}\ \emph {et~al.}(2018)\citenamefont
  {Ciarrocchi}, \citenamefont {Avsar},\ and\ \citenamefont
  {Ovchinnikov}}]{naturecomm2018}%
  \BibitemOpen
  \bibfield  {author} {\bibinfo {author} {\bibfnamefont {A.}~\bibnamefont
  {Ciarrocchi}}, \bibinfo {author} {\bibfnamefont {A.}~\bibnamefont {Avsar}},\
  and\ \bibinfo {author} {\bibfnamefont {D.~e.~a.}\ \bibnamefont
  {Ovchinnikov}},\ }\href {https://doi.org/10.1038/s41467-018-03436-0}
  {\bibfield  {journal} {\bibinfo  {journal} {Nat. Commun.}\ }\textbf {\bibinfo
  {volume} {9}},\ \bibinfo {pages} {919} (\bibinfo {year} {2018})}\BibitemShut
  {NoStop}%
\bibitem [{\citenamefont {Yan}\ \emph {et~al.}(2020)\citenamefont {Yan},
  \citenamefont {Wang}, \citenamefont {Lin}, \citenamefont {Wang},
  \citenamefont {Zeng}, \citenamefont {Boubeche}, \citenamefont {He},
  \citenamefont {Ma}, \citenamefont {Wang}, \citenamefont {Yao},\ and\
  \citenamefont {Luo}}]{NbSeTe}%
  \BibitemOpen
  \bibfield  {author} {\bibinfo {author} {\bibfnamefont {D.}~\bibnamefont
  {Yan}}, \bibinfo {author} {\bibfnamefont {S.}~\bibnamefont {Wang}}, \bibinfo
  {author} {\bibfnamefont {Y.}~\bibnamefont {Lin}}, \bibinfo {author}
  {\bibfnamefont {G.}~\bibnamefont {Wang}}, \bibinfo {author} {\bibfnamefont
  {Y.}~\bibnamefont {Zeng}}, \bibinfo {author} {\bibfnamefont {M.}~\bibnamefont
  {Boubeche}}, \bibinfo {author} {\bibfnamefont {Y.}~\bibnamefont {He}},
  \bibinfo {author} {\bibfnamefont {J.}~\bibnamefont {Ma}}, \bibinfo {author}
  {\bibfnamefont {Y.}~\bibnamefont {Wang}}, \bibinfo {author} {\bibfnamefont
  {D.-X.}\ \bibnamefont {Yao}},\ and\ \bibinfo {author} {\bibfnamefont
  {H.}~\bibnamefont {Luo}},\ }\href {https://doi.org/10.1088/1361-648X/ab46d0}
  {\bibfield  {journal} {\bibinfo  {journal} {Journal of Physics: Condensed
  Matter}\ }\textbf {\bibinfo {volume} {32}},\ \bibinfo {pages} {025702}
  (\bibinfo {year} {2020})}\BibitemShut {NoStop}%
\bibitem [{\citenamefont {Liu}\ \emph {et~al.}(2019)\citenamefont {Liu},
  \citenamefont {Gao}, \citenamefont {Zhang}, \citenamefont {He}, \citenamefont
  {Yu},\ and\ \citenamefont {Liu}}]{liu2019valleytronics}%
  \BibitemOpen
  \bibfield  {author} {\bibinfo {author} {\bibfnamefont {Y.}~\bibnamefont
  {Liu}}, \bibinfo {author} {\bibfnamefont {Y.}~\bibnamefont {Gao}}, \bibinfo
  {author} {\bibfnamefont {S.}~\bibnamefont {Zhang}}, \bibinfo {author}
  {\bibfnamefont {J.}~\bibnamefont {He}}, \bibinfo {author} {\bibfnamefont
  {J.}~\bibnamefont {Yu}},\ and\ \bibinfo {author} {\bibfnamefont
  {Z.}~\bibnamefont {Liu}},\ }\href {https://doi.org/10.1007/s12274-019-2497-2}
  {\bibfield  {journal} {\bibinfo  {journal} {Nano Research}\ }\textbf
  {\bibinfo {volume} {12}},\ \bibinfo {pages} {2695} (\bibinfo {year}
  {2019})}\BibitemShut {NoStop}%
\bibitem [{\citenamefont {Zibouche}\ \emph {et~al.}(2014)\citenamefont
  {Zibouche}, \citenamefont {Kuc}, \citenamefont {Musfeldt},\ and\
  \citenamefont {Heine}}]{zibouche2014transition}%
  \BibitemOpen
  \bibfield  {author} {\bibinfo {author} {\bibfnamefont {N.}~\bibnamefont
  {Zibouche}}, \bibinfo {author} {\bibfnamefont {A.}~\bibnamefont {Kuc}},
  \bibinfo {author} {\bibfnamefont {J.}~\bibnamefont {Musfeldt}},\ and\
  \bibinfo {author} {\bibfnamefont {T.}~\bibnamefont {Heine}},\ }\href
  {https://doi.org/10.1002/andp.201400137} {\bibfield  {journal} {\bibinfo
  {journal} {Annalen der Physik}\ }\textbf {\bibinfo {volume} {526}},\ \bibinfo
  {pages} {395} (\bibinfo {year} {2014})}\BibitemShut {NoStop}%
\bibitem [{\citenamefont {Mak}\ \emph {et~al.}(2014)\citenamefont {Mak},
  \citenamefont {McGill}, \citenamefont {Park},\ and\ \citenamefont
  {McEuen}}]{mak2014valley}%
  \BibitemOpen
  \bibfield  {author} {\bibinfo {author} {\bibfnamefont {K.~F.}\ \bibnamefont
  {Mak}}, \bibinfo {author} {\bibfnamefont {K.~L.}\ \bibnamefont {McGill}},
  \bibinfo {author} {\bibfnamefont {J.}~\bibnamefont {Park}},\ and\ \bibinfo
  {author} {\bibfnamefont {P.~L.}\ \bibnamefont {McEuen}},\ }\href
  {https://doi.org/10.1126/science.1250140} {\bibfield  {journal} {\bibinfo
  {journal} {Science}\ }\textbf {\bibinfo {volume} {344}},\ \bibinfo {pages}
  {1489} (\bibinfo {year} {2014})}\BibitemShut {NoStop}%
\bibitem [{\citenamefont {Xiao}\ \emph {et~al.}(2007)\citenamefont {Xiao},
  \citenamefont {Yao},\ and\ \citenamefont {Niu}}]{PhysRevLett.99.236809}%
  \BibitemOpen
  \bibfield  {author} {\bibinfo {author} {\bibfnamefont {D.}~\bibnamefont
  {Xiao}}, \bibinfo {author} {\bibfnamefont {W.}~\bibnamefont {Yao}},\ and\
  \bibinfo {author} {\bibfnamefont {Q.}~\bibnamefont {Niu}},\ }\href
  {https://doi.org/10.1103/PhysRevLett.99.236809} {\bibfield  {journal}
  {\bibinfo  {journal} {Phys. Rev. Lett.}\ }\textbf {\bibinfo {volume} {99}},\
  \bibinfo {pages} {236809} (\bibinfo {year} {2007})}\BibitemShut {NoStop}%
\bibitem [{\citenamefont {Bhowal}\ and\ \citenamefont
  {Satpathy}(2020)}]{Bhowal2020}%
  \BibitemOpen
  \bibfield  {author} {\bibinfo {author} {\bibfnamefont {S.}~\bibnamefont
  {Bhowal}}\ and\ \bibinfo {author} {\bibfnamefont {S.}~\bibnamefont
  {Satpathy}},\ }\href {https://doi.org/10.1103/PhysRevB.101.121112} {\bibfield
   {journal} {\bibinfo  {journal} {Phys. Rev. B}\ }\textbf {\bibinfo {volume}
  {101}},\ \bibinfo {pages} {121112(R)} (\bibinfo {year} {2020})}\BibitemShut
  {NoStop}%
\bibitem [{\citenamefont {Canonico}\ \emph {et~al.}(2020)\citenamefont
  {Canonico}, \citenamefont {Cysne}, \citenamefont {Molina-Sanchez},
  \citenamefont {Muniz},\ and\ \citenamefont {Rappoport}}]{Canonico2020}%
  \BibitemOpen
  \bibfield  {author} {\bibinfo {author} {\bibfnamefont {L.~M.}\ \bibnamefont
  {Canonico}}, \bibinfo {author} {\bibfnamefont {T.~P.}\ \bibnamefont {Cysne}},
  \bibinfo {author} {\bibfnamefont {A.}~\bibnamefont {Molina-Sanchez}},
  \bibinfo {author} {\bibfnamefont {R.~B.}\ \bibnamefont {Muniz}},\ and\
  \bibinfo {author} {\bibfnamefont {T.~G.}\ \bibnamefont {Rappoport}},\ }\href
  {https://doi.org/10.1103/PhysRevB.101.161409} {\bibfield  {journal} {\bibinfo
   {journal} {Phys. Rev. B}\ }\textbf {\bibinfo {volume} {101}},\ \bibinfo
  {pages} {161409(R)} (\bibinfo {year} {2020})}\BibitemShut {NoStop}%
\bibitem [{\citenamefont {Cysne}\ \emph {et~al.}(2021)\citenamefont {Cysne},
  \citenamefont {Costa}, \citenamefont {Canonico}, \citenamefont {Nardelli},
  \citenamefont {Muniz},\ and\ \citenamefont {Rappoport}}]{Cysne2021}%
  \BibitemOpen
  \bibfield  {author} {\bibinfo {author} {\bibfnamefont {T.~P.}\ \bibnamefont
  {Cysne}}, \bibinfo {author} {\bibfnamefont {M.}~\bibnamefont {Costa}},
  \bibinfo {author} {\bibfnamefont {L.~M.}\ \bibnamefont {Canonico}}, \bibinfo
  {author} {\bibfnamefont {M.~B.}\ \bibnamefont {Nardelli}}, \bibinfo {author}
  {\bibfnamefont {R.~B.}\ \bibnamefont {Muniz}},\ and\ \bibinfo {author}
  {\bibfnamefont {T.~G.}\ \bibnamefont {Rappoport}},\ }\href
  {https://doi.org/10.1103/PhysRevLett.126.056601} {\bibfield  {journal}
  {\bibinfo  {journal} {Phys. Rev. Lett.}\ }\textbf {\bibinfo {volume} {126}},\
  \bibinfo {pages} {056601} (\bibinfo {year} {2021})}\BibitemShut {NoStop}%
\bibitem [{\citenamefont {Bhowal}\ and\ \citenamefont
  {Vignale}(2021)}]{Vignale2021}%
  \BibitemOpen
  \bibfield  {author} {\bibinfo {author} {\bibfnamefont {S.}~\bibnamefont
  {Bhowal}}\ and\ \bibinfo {author} {\bibfnamefont {G.}~\bibnamefont
  {Vignale}},\ }\href {https://doi.org/10.1103/PhysRevB.103.195309} {\bibfield
  {journal} {\bibinfo  {journal} {Phys. Rev. B}\ }\textbf {\bibinfo {volume}
  {103}},\ \bibinfo {pages} {195309} (\bibinfo {year} {2021})}\BibitemShut
  {NoStop}%
\bibitem [{\citenamefont {Xiao}\ \emph {et~al.}(2012)\citenamefont {Xiao},
  \citenamefont {Liu}, \citenamefont {Feng}, \citenamefont {Xu},\ and\
  \citenamefont {Yao}}]{PhysRevLett.108.196802}%
  \BibitemOpen
  \bibfield  {author} {\bibinfo {author} {\bibfnamefont {D.}~\bibnamefont
  {Xiao}}, \bibinfo {author} {\bibfnamefont {G.-B.}\ \bibnamefont {Liu}},
  \bibinfo {author} {\bibfnamefont {W.}~\bibnamefont {Feng}}, \bibinfo {author}
  {\bibfnamefont {X.}~\bibnamefont {Xu}},\ and\ \bibinfo {author}
  {\bibfnamefont {W.}~\bibnamefont {Yao}},\ }\href
  {https://doi.org/10.1103/PhysRevLett.108.196802} {\bibfield  {journal}
  {\bibinfo  {journal} {Phys. Rev. Lett.}\ }\textbf {\bibinfo {volume} {108}},\
  \bibinfo {pages} {196802} (\bibinfo {year} {2012})}\BibitemShut {NoStop}%
\bibitem [{\citenamefont {Ma}\ \emph {et~al.}(2019)\citenamefont {Ma},
  \citenamefont {Xu}, \citenamefont {Shen} \emph {et~al.}}]{Nature2019}%
  \BibitemOpen
  \bibfield  {author} {\bibinfo {author} {\bibfnamefont {Q.}~\bibnamefont
  {Ma}}, \bibinfo {author} {\bibfnamefont {S.}~\bibnamefont {Xu}}, \bibinfo
  {author} {\bibfnamefont {H.}~\bibnamefont {Shen}}, \emph {et~al.},\ }\href
  {https://doi.org/10.1038/s41586-018-0807-6} {\bibfield  {journal} {\bibinfo
  {journal} {Nature}\ }\textbf {\bibinfo {volume} {565}},\ \bibinfo {pages}
  {337} (\bibinfo {year} {2019})}\BibitemShut {NoStop}%
\bibitem [{\citenamefont {Sodemann}\ and\ \citenamefont
  {Fu}(2015)}]{Liang2015}%
  \BibitemOpen
  \bibfield  {author} {\bibinfo {author} {\bibfnamefont {I.}~\bibnamefont
  {Sodemann}}\ and\ \bibinfo {author} {\bibfnamefont {L.}~\bibnamefont {Fu}},\
  }\href {https://doi.org/10.1103/PhysRevLett.115.216806} {\bibfield  {journal}
  {\bibinfo  {journal} {Phys. Rev. Lett.}\ }\textbf {\bibinfo {volume} {115}},\
  \bibinfo {pages} {216806} (\bibinfo {year} {2015})}\BibitemShut {NoStop}%
\bibitem [{\citenamefont {Du}\ \emph {et~al.}(2018)\citenamefont {Du},
  \citenamefont {Wang}, \citenamefont {Lu},\ and\ \citenamefont
  {Xie}}]{Xie2018}%
  \BibitemOpen
  \bibfield  {author} {\bibinfo {author} {\bibfnamefont {Z.~Z.}\ \bibnamefont
  {Du}}, \bibinfo {author} {\bibfnamefont {C.~M.}\ \bibnamefont {Wang}},
  \bibinfo {author} {\bibfnamefont {H.-Z.}\ \bibnamefont {Lu}},\ and\ \bibinfo
  {author} {\bibfnamefont {X.~C.}\ \bibnamefont {Xie}},\ }\href
  {https://doi.org/10.1103/PhysRevLett.121.266601} {\bibfield  {journal}
  {\bibinfo  {journal} {Phys. Rev. Lett.}\ }\textbf {\bibinfo {volume} {121}},\
  \bibinfo {pages} {266601} (\bibinfo {year} {2018})}\BibitemShut {NoStop}%
\bibitem [{\citenamefont {Qin}\ \emph {et~al.}(2021)\citenamefont {Qin},
  \citenamefont {Zhu}, \citenamefont {Ye}, \citenamefont {Xu}, \citenamefont
  {Song}, \citenamefont {Liang}, \citenamefont {Liu},\ and\ \citenamefont
  {Liao}}]{Qin_2021}%
  \BibitemOpen
  \bibfield  {author} {\bibinfo {author} {\bibfnamefont {M.-S.}\ \bibnamefont
  {Qin}}, \bibinfo {author} {\bibfnamefont {P.-F.}\ \bibnamefont {Zhu}},
  \bibinfo {author} {\bibfnamefont {X.-G.}\ \bibnamefont {Ye}}, \bibinfo
  {author} {\bibfnamefont {W.-Z.}\ \bibnamefont {Xu}}, \bibinfo {author}
  {\bibfnamefont {Z.-H.}\ \bibnamefont {Song}}, \bibinfo {author}
  {\bibfnamefont {J.}~\bibnamefont {Liang}}, \bibinfo {author} {\bibfnamefont
  {K.}~\bibnamefont {Liu}},\ and\ \bibinfo {author} {\bibfnamefont {Z.-M.}\
  \bibnamefont {Liao}},\ }\href {https://doi.org/10.1088/0256-307x/38/1/017301}
  {\bibfield  {journal} {\bibinfo  {journal} {Chinese Physics Letters}\
  }\textbf {\bibinfo {volume} {38}},\ \bibinfo {pages} {017301} (\bibinfo
  {year} {2021})}\BibitemShut {NoStop}%
\bibitem [{\citenamefont {Vargiamidis}\ \emph {et~al.}(2020)\citenamefont
  {Vargiamidis}, \citenamefont {Vasilopoulos}, \citenamefont {Tahir},\ and\
  \citenamefont {Neophytou}}]{PRB2020}%
  \BibitemOpen
  \bibfield  {author} {\bibinfo {author} {\bibfnamefont {V.}~\bibnamefont
  {Vargiamidis}}, \bibinfo {author} {\bibfnamefont {P.}~\bibnamefont
  {Vasilopoulos}}, \bibinfo {author} {\bibfnamefont {M.}~\bibnamefont
  {Tahir}},\ and\ \bibinfo {author} {\bibfnamefont {N.}~\bibnamefont
  {Neophytou}},\ }\href {https://doi.org/10.1103/PhysRevB.102.235426}
  {\bibfield  {journal} {\bibinfo  {journal} {Phys. Rev. B}\ }\textbf {\bibinfo
  {volume} {102}},\ \bibinfo {pages} {235426} (\bibinfo {year}
  {2020})}\BibitemShut {NoStop}%
\bibitem [{\citenamefont {Zhou}\ \emph {et~al.}(2019)\citenamefont {Zhou},
  \citenamefont {Taguchi}, \citenamefont {Kawaguchi}, \citenamefont {Yukio},\
  and\ \citenamefont {Law}}]{CommPhys2019}%
  \BibitemOpen
  \bibfield  {author} {\bibinfo {author} {\bibfnamefont {B.~T.}\ \bibnamefont
  {Zhou}}, \bibinfo {author} {\bibfnamefont {K.}~\bibnamefont {Taguchi}},
  \bibinfo {author} {\bibfnamefont {Y.}~\bibnamefont {Kawaguchi}}, \bibinfo
  {author} {\bibfnamefont {T.}~\bibnamefont {Yukio}},\ and\ \bibinfo {author}
  {\bibfnamefont {K.}~\bibnamefont {Law}},\ }\href
  {https://doi.org/10.1038/s42005-019-0127-7} {\bibfield  {journal} {\bibinfo
  {journal} {Commun. Phys.}\ }\textbf {\bibinfo {volume} {2}},\ \bibinfo
  {pages} {26} (\bibinfo {year} {2019})}\BibitemShut {NoStop}%
\bibitem [{\citenamefont {Zhao}\ \emph {et~al.}(2017)\citenamefont {Zhao},
  \citenamefont {Norden}, \citenamefont {Zhang}, \citenamefont {Zhao},
  \citenamefont {Cheng}, \citenamefont {Sun}, \citenamefont {Parry},
  \citenamefont {Taheri}, \citenamefont {Wang}, \citenamefont {Yang} \emph
  {et~al.}}]{zhao2017enhanced}%
  \BibitemOpen
  \bibfield  {author} {\bibinfo {author} {\bibfnamefont {C.}~\bibnamefont
  {Zhao}}, \bibinfo {author} {\bibfnamefont {T.}~\bibnamefont {Norden}},
  \bibinfo {author} {\bibfnamefont {P.}~\bibnamefont {Zhang}}, \bibinfo
  {author} {\bibfnamefont {P.}~\bibnamefont {Zhao}}, \bibinfo {author}
  {\bibfnamefont {Y.}~\bibnamefont {Cheng}}, \bibinfo {author} {\bibfnamefont
  {F.}~\bibnamefont {Sun}}, \bibinfo {author} {\bibfnamefont {J.~P.}\
  \bibnamefont {Parry}}, \bibinfo {author} {\bibfnamefont {P.}~\bibnamefont
  {Taheri}}, \bibinfo {author} {\bibfnamefont {J.}~\bibnamefont {Wang}},
  \bibinfo {author} {\bibfnamefont {Y.}~\bibnamefont {Yang}}, \emph {et~al.},\
  }\href {https://doi.org/10.1038/nnano.2017.68} {\bibfield  {journal}
  {\bibinfo  {journal} {Nature nanotechnology}\ }\textbf {\bibinfo {volume}
  {12}},\ \bibinfo {pages} {757} (\bibinfo {year} {2017})}\BibitemShut
  {NoStop}%
\bibitem [{\citenamefont {Da}\ \emph {et~al.}(2020)\citenamefont {Da},
  \citenamefont {Song}, \citenamefont {Dong}, \citenamefont {Ye},\ and\
  \citenamefont {Yan}}]{JAP2020}%
  \BibitemOpen
  \bibfield  {author} {\bibinfo {author} {\bibfnamefont {H.}~\bibnamefont
  {Da}}, \bibinfo {author} {\bibfnamefont {Q.}~\bibnamefont {Song}}, \bibinfo
  {author} {\bibfnamefont {P.}~\bibnamefont {Dong}}, \bibinfo {author}
  {\bibfnamefont {H.}~\bibnamefont {Ye}},\ and\ \bibinfo {author}
  {\bibfnamefont {X.}~\bibnamefont {Yan}},\ }\href
  {https://doi.org/10.1063/1.5118327} {\bibfield  {journal} {\bibinfo
  {journal} {Journal of Applied Physics}\ }\textbf {\bibinfo {volume} {127}},\
  \bibinfo {pages} {023903} (\bibinfo {year} {2020})}\BibitemShut {NoStop}%
\bibitem [{\citenamefont {Habe}\ and\ \citenamefont {Koshino}(2017)}]{PRB2017}%
  \BibitemOpen
  \bibfield  {author} {\bibinfo {author} {\bibfnamefont {T.}~\bibnamefont
  {Habe}}\ and\ \bibinfo {author} {\bibfnamefont {M.}~\bibnamefont {Koshino}},\
  }\href {https://doi.org/10.1103/PhysRevB.96.085411} {\bibfield  {journal}
  {\bibinfo  {journal} {Phys. Rev. B}\ }\textbf {\bibinfo {volume} {96}},\
  \bibinfo {pages} {085411} (\bibinfo {year} {2017})}\BibitemShut {NoStop}%
\bibitem [{\citenamefont {Hidding}\ and\ \citenamefont
  {Guimarães}(2020)}]{Hidding2020}%
  \BibitemOpen
  \bibfield  {author} {\bibinfo {author} {\bibfnamefont {J.}~\bibnamefont
  {Hidding}}\ and\ \bibinfo {author} {\bibfnamefont {M.~H.~D.}\ \bibnamefont
  {Guimarães}},\ }\href {https://doi.org/10.3389/fmats.2020.594771} {\bibfield
   {journal} {\bibinfo  {journal} {Frontiers in Materials}\ }\textbf {\bibinfo
  {volume} {7}},\ \bibinfo {pages} {383} (\bibinfo {year} {2020})}\BibitemShut
  {NoStop}%
\bibitem [{\citenamefont {Gambardella}\ and\ \citenamefont
  {Miron}(2011)}]{Gambardella2011}%
  \BibitemOpen
  \bibfield  {author} {\bibinfo {author} {\bibfnamefont {P.}~\bibnamefont
  {Gambardella}}\ and\ \bibinfo {author} {\bibfnamefont {I.~M.}\ \bibnamefont
  {Miron}},\ }\bibfield  {journal} {\bibinfo  {journal} {Philos. Trans. R. Soc.
  A Math. Phys. Eng. Sci.}\ }\textbf {\bibinfo {volume} {369}},\ \href
  {https://doi.org/10.1098/rsta.2010.0336} {10.1098/rsta.2010.0336} (\bibinfo
  {year} {2011})\BibitemShut {NoStop}%
\bibitem [{\citenamefont {MacNeill}\ \emph {et~al.}(2017)\citenamefont
  {MacNeill}, \citenamefont {Stiehl}, \citenamefont {Guimaraes}, \citenamefont
  {Buhrman}, \citenamefont {Park},\ and\ \citenamefont {Ralph}}]{MacNeill2017}%
  \BibitemOpen
  \bibfield  {author} {\bibinfo {author} {\bibfnamefont {D.}~\bibnamefont
  {MacNeill}}, \bibinfo {author} {\bibfnamefont {G.~M.}\ \bibnamefont
  {Stiehl}}, \bibinfo {author} {\bibfnamefont {M.~H.~D.}\ \bibnamefont
  {Guimaraes}}, \bibinfo {author} {\bibfnamefont {R.~A.}\ \bibnamefont
  {Buhrman}}, \bibinfo {author} {\bibfnamefont {J.}~\bibnamefont {Park}},\ and\
  \bibinfo {author} {\bibfnamefont {D.~C.}\ \bibnamefont {Ralph}},\ }\href
  {https://doi.org/10.1038/nphys3933} {\bibfield  {journal} {\bibinfo
  {journal} {Nature Physics}\ }\textbf {\bibinfo {volume} {13}},\ \bibinfo
  {pages} {300} (\bibinfo {year} {2017})}\BibitemShut {NoStop}%
\bibitem [{\citenamefont {Xue}\ \emph {et~al.}(2020)\citenamefont {Xue},
  \citenamefont {Rohmann}, \citenamefont {Li}, \citenamefont {Amin},\ and\
  \citenamefont {Haney}}]{Xue2020}%
  \BibitemOpen
  \bibfield  {author} {\bibinfo {author} {\bibfnamefont {F.}~\bibnamefont
  {Xue}}, \bibinfo {author} {\bibfnamefont {C.}~\bibnamefont {Rohmann}},
  \bibinfo {author} {\bibfnamefont {J.}~\bibnamefont {Li}}, \bibinfo {author}
  {\bibfnamefont {V.}~\bibnamefont {Amin}},\ and\ \bibinfo {author}
  {\bibfnamefont {P.}~\bibnamefont {Haney}},\ }\href
  {https://doi.org/10.1103/PhysRevB.102.014401} {\bibfield  {journal} {\bibinfo
   {journal} {Phys. Rev. B}\ }\textbf {\bibinfo {volume} {102}},\ \bibinfo
  {pages} {014401} (\bibinfo {year} {2020})}\BibitemShut {NoStop}%
\bibitem [{\citenamefont {Miyamachi}\ \emph {et~al.}(2013)\citenamefont
  {Miyamachi}, \citenamefont {Schuh}, \citenamefont {Märkl}, \citenamefont
  {Bresch}, \citenamefont {Balashov}, \citenamefont {Stöhr}, \citenamefont
  {Karlewski}, \citenamefont {André}, \citenamefont {Marthaler}, \citenamefont
  {Hoffmann}, \citenamefont {Geilhufe}, \citenamefont {Ostanin}, \citenamefont
  {Hergert}, \citenamefont {Mertig}, \citenamefont {Schön}, \citenamefont
  {Ernst},\ and\ \citenamefont {Wulfhekel}}]{Nature2013}%
  \BibitemOpen
  \bibfield  {author} {\bibinfo {author} {\bibfnamefont {T.}~\bibnamefont
  {Miyamachi}}, \bibinfo {author} {\bibfnamefont {T.}~\bibnamefont {Schuh}},
  \bibinfo {author} {\bibfnamefont {T.}~\bibnamefont {Märkl}}, \bibinfo
  {author} {\bibfnamefont {C.}~\bibnamefont {Bresch}}, \bibinfo {author}
  {\bibfnamefont {T.}~\bibnamefont {Balashov}}, \bibinfo {author}
  {\bibfnamefont {A.}~\bibnamefont {Stöhr}}, \bibinfo {author} {\bibfnamefont
  {C.}~\bibnamefont {Karlewski}}, \bibinfo {author} {\bibfnamefont
  {S.}~\bibnamefont {André}}, \bibinfo {author} {\bibfnamefont
  {M.}~\bibnamefont {Marthaler}}, \bibinfo {author} {\bibfnamefont
  {M.}~\bibnamefont {Hoffmann}}, \bibinfo {author} {\bibfnamefont
  {M.}~\bibnamefont {Geilhufe}}, \bibinfo {author} {\bibfnamefont
  {S.}~\bibnamefont {Ostanin}}, \bibinfo {author} {\bibfnamefont
  {W.}~\bibnamefont {Hergert}}, \bibinfo {author} {\bibfnamefont
  {I.}~\bibnamefont {Mertig}}, \bibinfo {author} {\bibfnamefont
  {G.}~\bibnamefont {Schön}}, \bibinfo {author} {\bibfnamefont
  {A.}~\bibnamefont {Ernst}},\ and\ \bibinfo {author} {\bibfnamefont
  {W.}~\bibnamefont {Wulfhekel}},\ }\href {https://doi.org/10.1038/nature12759}
  {\bibfield  {journal} {\bibinfo  {journal} {Nature}\ }\textbf {\bibinfo
  {volume} {503}},\ \bibinfo {pages} {242} (\bibinfo {year}
  {2013})}\BibitemShut {NoStop}%
\bibitem [{\citenamefont {Donati}\ \emph {et~al.}(2016)\citenamefont {Donati},
  \citenamefont {Rusponi}, \citenamefont {Stepanow}, \citenamefont
  {W{\"a}ckerlin}, \citenamefont {Singha}, \citenamefont {Persichetti},
  \citenamefont {Baltic}, \citenamefont {Diller}, \citenamefont {Patthey},
  \citenamefont {Fernandes}, \citenamefont {Dreiser}, \citenamefont {{\v
  S}ljivan{\v c}anin}, \citenamefont {Kummer}, \citenamefont {Nistor},
  \citenamefont {Gambardella},\ and\ \citenamefont {Brune}}]{Science2016}%
  \BibitemOpen
  \bibfield  {author} {\bibinfo {author} {\bibfnamefont {F.}~\bibnamefont
  {Donati}}, \bibinfo {author} {\bibfnamefont {S.}~\bibnamefont {Rusponi}},
  \bibinfo {author} {\bibfnamefont {S.}~\bibnamefont {Stepanow}}, \bibinfo
  {author} {\bibfnamefont {C.}~\bibnamefont {W{\"a}ckerlin}}, \bibinfo {author}
  {\bibfnamefont {A.}~\bibnamefont {Singha}}, \bibinfo {author} {\bibfnamefont
  {L.}~\bibnamefont {Persichetti}}, \bibinfo {author} {\bibfnamefont
  {R.}~\bibnamefont {Baltic}}, \bibinfo {author} {\bibfnamefont
  {K.}~\bibnamefont {Diller}}, \bibinfo {author} {\bibfnamefont
  {F.}~\bibnamefont {Patthey}}, \bibinfo {author} {\bibfnamefont
  {E.}~\bibnamefont {Fernandes}}, \bibinfo {author} {\bibfnamefont
  {J.}~\bibnamefont {Dreiser}}, \bibinfo {author} {\bibfnamefont {{\v
  Z}.}~\bibnamefont {{\v S}ljivan{\v c}anin}}, \bibinfo {author} {\bibfnamefont
  {K.}~\bibnamefont {Kummer}}, \bibinfo {author} {\bibfnamefont
  {C.}~\bibnamefont {Nistor}}, \bibinfo {author} {\bibfnamefont
  {P.}~\bibnamefont {Gambardella}},\ and\ \bibinfo {author} {\bibfnamefont
  {H.}~\bibnamefont {Brune}},\ }\href {https://doi.org/10.1126/science.aad9898}
  {\bibfield  {journal} {\bibinfo  {journal} {Science}\ }\textbf {\bibinfo
  {volume} {352}},\ \bibinfo {pages} {318} (\bibinfo {year}
  {2016})}\BibitemShut {NoStop}%
\bibitem [{\citenamefont {Baltic}\ \emph {et~al.}(2016)\citenamefont {Baltic},
  \citenamefont {Pivetta}, \citenamefont {Donati}, \citenamefont {Wäckerlin},
  \citenamefont {Singha}, \citenamefont {Dreiser}, \citenamefont {Rusponi},\
  and\ \citenamefont {Brune}}]{Nano2016}%
  \BibitemOpen
  \bibfield  {author} {\bibinfo {author} {\bibfnamefont {R.}~\bibnamefont
  {Baltic}}, \bibinfo {author} {\bibfnamefont {M.}~\bibnamefont {Pivetta}},
  \bibinfo {author} {\bibfnamefont {F.}~\bibnamefont {Donati}}, \bibinfo
  {author} {\bibfnamefont {C.}~\bibnamefont {Wäckerlin}}, \bibinfo {author}
  {\bibfnamefont {A.}~\bibnamefont {Singha}}, \bibinfo {author} {\bibfnamefont
  {J.}~\bibnamefont {Dreiser}}, \bibinfo {author} {\bibfnamefont
  {S.}~\bibnamefont {Rusponi}},\ and\ \bibinfo {author} {\bibfnamefont
  {H.}~\bibnamefont {Brune}},\ }\href
  {https://doi.org/10.1021/acs.nanolett.6b03543} {\bibfield  {journal}
  {\bibinfo  {journal} {Nano Letters}\ }\textbf {\bibinfo {volume} {16}},\
  \bibinfo {pages} {7610} (\bibinfo {year} {2016})}\BibitemShut {NoStop}%
\bibitem [{\citenamefont {Shick}\ and\ \citenamefont
  {Lichtenstein}(2018)}]{Shick2018}%
  \BibitemOpen
  \bibfield  {author} {\bibinfo {author} {\bibfnamefont {A.}~\bibnamefont
  {Shick}}\ and\ \bibinfo {author} {\bibfnamefont {A.}~\bibnamefont
  {Lichtenstein}},\ }\href
  {https://doi.org/https://doi.org/10.1016/j.jmmm.2018.01.050} {\bibfield
  {journal} {\bibinfo  {journal} {Journal of Magnetism and Magnetic Materials}\
  }\textbf {\bibinfo {volume} {454}},\ \bibinfo {pages} {61} (\bibinfo {year}
  {2018})}\BibitemShut {NoStop}%
\bibitem [{\citenamefont {Shick}\ and\ \citenamefont
  {Denisov}(2019)}]{Shick2019}%
  \BibitemOpen
  \bibfield  {author} {\bibinfo {author} {\bibfnamefont {A.}~\bibnamefont
  {Shick}}\ and\ \bibinfo {author} {\bibfnamefont {A.}~\bibnamefont
  {Denisov}},\ }\href
  {https://doi.org/https://doi.org/10.1016/j.jmmm.2018.11.078} {\bibfield
  {journal} {\bibinfo  {journal} {Journal of Magnetism and Magnetic Materials}\
  }\textbf {\bibinfo {volume} {475}},\ \bibinfo {pages} {211} (\bibinfo {year}
  {2019})}\BibitemShut {NoStop}%
\bibitem [{FLE()}]{FLEUR}%
  \BibitemOpen
  \href@noop {} {}\bibinfo {howpublished}
  {\url{https://www.flapw.de/}}\BibitemShut {NoStop}%
\bibitem [{\citenamefont {Shick}\ \emph {et~al.}(1999)\citenamefont {Shick},
  \citenamefont {Liechtenstein},\ and\ \citenamefont {Pickett}}]{DFTUShick}%
  \BibitemOpen
  \bibfield  {author} {\bibinfo {author} {\bibfnamefont {A.~B.}\ \bibnamefont
  {Shick}}, \bibinfo {author} {\bibfnamefont {A.~I.}\ \bibnamefont
  {Liechtenstein}},\ and\ \bibinfo {author} {\bibfnamefont {W.~E.}\
  \bibnamefont {Pickett}},\ }\href {https://doi.org/10.1103/PhysRevB.60.10763}
  {\bibfield  {journal} {\bibinfo  {journal} {Phys. Rev. B}\ }\textbf {\bibinfo
  {volume} {60}},\ \bibinfo {pages} {10763} (\bibinfo {year}
  {1999})}\BibitemShut {NoStop}%
\bibitem [{\citenamefont {Perdew}\ \emph {et~al.}(1996)\citenamefont {Perdew},
  \citenamefont {Burke},\ and\ \citenamefont {Ernzerhof}}]{PBE}%
  \BibitemOpen
  \bibfield  {author} {\bibinfo {author} {\bibfnamefont {J.~P.}\ \bibnamefont
  {Perdew}}, \bibinfo {author} {\bibfnamefont {K.}~\bibnamefont {Burke}},\ and\
  \bibinfo {author} {\bibfnamefont {M.}~\bibnamefont {Ernzerhof}},\ }\href
  {https://doi.org/10.1103/PhysRevLett.77.3865} {\bibfield  {journal} {\bibinfo
   {journal} {Phys. Rev. Lett.}\ }\textbf {\bibinfo {volume} {77}},\ \bibinfo
  {pages} {3865} (\bibinfo {year} {1996})}\BibitemShut {NoStop}%
\bibitem [{\citenamefont {Kurz}\ \emph {et~al.}(2002)\citenamefont {Kurz},
  \citenamefont {Bihlmayer},\ and\ \citenamefont {Blügel}}]{Kurz_2002}%
  \BibitemOpen
  \bibfield  {author} {\bibinfo {author} {\bibfnamefont {P.}~\bibnamefont
  {Kurz}}, \bibinfo {author} {\bibfnamefont {G.}~\bibnamefont {Bihlmayer}},\
  and\ \bibinfo {author} {\bibfnamefont {S.}~\bibnamefont {Blügel}},\ }\href
  {https://doi.org/10.1088/0953-8984/14/25/305} {\bibfield  {journal} {\bibinfo
   {journal} {Journal of Physics: Condensed Matter}\ }\textbf {\bibinfo
  {volume} {14}},\ \bibinfo {pages} {6353} (\bibinfo {year}
  {2002})}\BibitemShut {NoStop}%
\bibitem [{\citenamefont {Freimuth}\ \emph {et~al.}(2008)\citenamefont
  {Freimuth}, \citenamefont {Mokrousov}, \citenamefont {Wortmann},
  \citenamefont {Heinze},\ and\ \citenamefont {Blügel}}]{Freimuth_2008}%
  \BibitemOpen
  \bibfield  {author} {\bibinfo {author} {\bibfnamefont {F.}~\bibnamefont
  {Freimuth}}, \bibinfo {author} {\bibfnamefont {Y.}~\bibnamefont {Mokrousov}},
  \bibinfo {author} {\bibfnamefont {D.}~\bibnamefont {Wortmann}}, \bibinfo
  {author} {\bibfnamefont {S.}~\bibnamefont {Heinze}},\ and\ \bibinfo {author}
  {\bibfnamefont {S.}~\bibnamefont {Blügel}},\ }\bibfield  {journal} {\bibinfo
   {journal} {Physical Review B}\ }\textbf {\bibinfo {volume} {78}},\ \href
  {https://doi.org/10.1103/physrevb.78.035120} {10.1103/physrevb.78.035120}
  (\bibinfo {year} {2008})\BibitemShut {NoStop}%
\bibitem [{\citenamefont {Mostofi}\ \emph {et~al.}(2014)\citenamefont
  {Mostofi}, \citenamefont {Yates}, \citenamefont {Pizzi}, \citenamefont {Lee},
  \citenamefont {Souza}, \citenamefont {Vanderbilt},\ and\ \citenamefont
  {Marzari}}]{Wannier90}%
  \BibitemOpen
  \bibfield  {author} {\bibinfo {author} {\bibfnamefont {A.~A.}\ \bibnamefont
  {Mostofi}}, \bibinfo {author} {\bibfnamefont {J.~R.}\ \bibnamefont {Yates}},
  \bibinfo {author} {\bibfnamefont {G.}~\bibnamefont {Pizzi}}, \bibinfo
  {author} {\bibfnamefont {Y.-S.}\ \bibnamefont {Lee}}, \bibinfo {author}
  {\bibfnamefont {I.}~\bibnamefont {Souza}}, \bibinfo {author} {\bibfnamefont
  {D.}~\bibnamefont {Vanderbilt}},\ and\ \bibinfo {author} {\bibfnamefont
  {N.}~\bibnamefont {Marzari}},\ }\href
  {https://doi.org/https://doi.org/10.1016/j.cpc.2014.05.003} {\bibfield
  {journal} {\bibinfo  {journal} {Computer Physics Communications}\ }\textbf
  {\bibinfo {volume} {185}},\ \bibinfo {pages} {2309} (\bibinfo {year}
  {2014})}\BibitemShut {NoStop}%
\bibitem [{\citenamefont {Mermin}\ and\ \citenamefont
  {Wagner}(1966)}]{Mermin1966}%
  \BibitemOpen
  \bibfield  {author} {\bibinfo {author} {\bibfnamefont {N.~D.}\ \bibnamefont
  {Mermin}}\ and\ \bibinfo {author} {\bibfnamefont {H.}~\bibnamefont
  {Wagner}},\ }\href {https://doi.org/10.1103/PhysRevLett.17.1133} {\bibfield
  {journal} {\bibinfo  {journal} {Phys. Rev. Lett.}\ }\textbf {\bibinfo
  {volume} {17}},\ \bibinfo {pages} {1133} (\bibinfo {year}
  {1966})}\BibitemShut {NoStop}%
\bibitem [{\citenamefont {Griffiths}(1964)}]{Griffiths1964}%
  \BibitemOpen
  \bibfield  {author} {\bibinfo {author} {\bibfnamefont {R.~B.}\ \bibnamefont
  {Griffiths}},\ }\href {https://doi.org/10.1103/PhysRev.136.A437} {\bibfield
  {journal} {\bibinfo  {journal} {Phys. Rev.}\ }\textbf {\bibinfo {volume}
  {136}},\ \bibinfo {pages} {437} (\bibinfo {year} {1964})}\BibitemShut
  {NoStop}%
\bibitem [{\citenamefont {Kurz}\ \emph {et~al.}(2004)\citenamefont {Kurz},
  \citenamefont {F\"orster}, \citenamefont {Nordstr\"om}, \citenamefont
  {Bihlmayer},\ and\ \citenamefont {Bl\"ugel}}]{Kurz2004}%
  \BibitemOpen
  \bibfield  {author} {\bibinfo {author} {\bibfnamefont {P.}~\bibnamefont
  {Kurz}}, \bibinfo {author} {\bibfnamefont {F.}~\bibnamefont {F\"orster}},
  \bibinfo {author} {\bibfnamefont {L.}~\bibnamefont {Nordstr\"om}}, \bibinfo
  {author} {\bibfnamefont {G.}~\bibnamefont {Bihlmayer}},\ and\ \bibinfo
  {author} {\bibfnamefont {S.}~\bibnamefont {Bl\"ugel}},\ }\href
  {https://doi.org/10.1103/PhysRevB.69.024415} {\bibfield  {journal} {\bibinfo
  {journal} {Phys. Rev. B}\ }\textbf {\bibinfo {volume} {69}},\ \bibinfo
  {pages} {024415} (\bibinfo {year} {2004})}\BibitemShut {NoStop}%
\bibitem [{\citenamefont {Heide}\ \emph {et~al.}(2009)\citenamefont {Heide},
  \citenamefont {Bihlmayer},\ and\ \citenamefont {Blügel}}]{Heide2009}%
  \BibitemOpen
  \bibfield  {author} {\bibinfo {author} {\bibfnamefont {M.}~\bibnamefont
  {Heide}}, \bibinfo {author} {\bibfnamefont {G.}~\bibnamefont {Bihlmayer}},\
  and\ \bibinfo {author} {\bibfnamefont {S.}~\bibnamefont {Blügel}},\ }\href
  {https://doi.org/https://doi.org/10.1016/j.physb.2009.06.070} {\bibfield
  {journal} {\bibinfo  {journal} {Physica B: Condensed Matter}\ }\textbf
  {\bibinfo {volume} {404}},\ \bibinfo {pages} {2678} (\bibinfo {year}
  {2009})}\BibitemShut {NoStop}%
\bibitem [{\citenamefont {Rado}\ and\ \citenamefont
  {Suhl}(1966)}]{rado1966magnetism}%
  \BibitemOpen
  \bibfield  {author} {\bibinfo {author} {\bibfnamefont {G.}~\bibnamefont
  {Rado}}\ and\ \bibinfo {author} {\bibfnamefont {H.}~\bibnamefont {Suhl}},\
  }\href {https://books.google.de/books?id=kcLvAAAAMAAJ} {}Magnetism\ (\bibinfo
   {publisher} {Academic Press},\ \bibinfo {year} {1966})\BibitemShut {NoStop}%
\bibitem [{\citenamefont {Sandratskii}(1986)}]{Sandratskii1986}%
  \BibitemOpen
  \bibfield  {author} {\bibinfo {author} {\bibfnamefont {L.~M.}\ \bibnamefont
  {Sandratskii}},\ }\href
  {https://doi.org/https://doi.org/10.1002/pssb.2221360119} {\bibfield
  {journal} {\bibinfo  {journal} {physica status solidi (b)}\ }\textbf
  {\bibinfo {volume} {136}},\ \bibinfo {pages} {167} (\bibinfo {year}
  {1986})}\BibitemShut {NoStop}%
\bibitem [{\citenamefont {{Bychkov}}\ and\ \citenamefont
  {{Rashba}}(1984)}]{Rashba1984}%
  \BibitemOpen
  \bibfield  {author} {\bibinfo {author} {\bibfnamefont {Y.~A.}\ \bibnamefont
  {{Bychkov}}}\ and\ \bibinfo {author} {\bibfnamefont {{\'E}.~I.}\ \bibnamefont
  {{Rashba}}},\ }\href@noop {} {\bibfield  {journal} {\bibinfo  {journal}
  {Soviet Journal of Experimental and Theoretical Physics Letters}\ }\textbf
  {\bibinfo {volume} {39}},\ \bibinfo {pages} {78} (\bibinfo {year}
  {1984})}\BibitemShut {NoStop}%
\bibitem [{\citenamefont {Manchon}\ \emph {et~al.}(2015)\citenamefont
  {Manchon}, \citenamefont {Koo}, \citenamefont {Nitta}, \citenamefont
  {Frolov},\ and\ \citenamefont {Duine}}]{Manchon2015}%
  \BibitemOpen
  \bibfield  {author} {\bibinfo {author} {\bibfnamefont {A.}~\bibnamefont
  {Manchon}}, \bibinfo {author} {\bibfnamefont {H.~C.}\ \bibnamefont {Koo}},
  \bibinfo {author} {\bibfnamefont {J.}~\bibnamefont {Nitta}}, \bibinfo
  {author} {\bibfnamefont {S.~M.}\ \bibnamefont {Frolov}},\ and\ \bibinfo
  {author} {\bibfnamefont {R.~A.}\ \bibnamefont {Duine}},\ }\bibfield  {title}
  {\bibinfo {title} {{New perspectives for Rashba spin--orbit coupling}},\
  }\href {https://doi.org/10.1038/nmat4360} {\bibfield  {journal} {\bibinfo
  {journal} {Nature Materials}\ }\textbf {\bibinfo {volume} {14}},\ \bibinfo
  {pages} {871} (\bibinfo {year} {2015})}\BibitemShut {NoStop}%
\bibitem [{Par(2011)}]{Park2011}%
  \BibitemOpen
  \href {https://doi.org/10.1103/PhysRevLett.107.156803} {\bibfield  {journal}
  {\bibinfo  {journal} {Phys. Rev. Lett.}\ }\textbf {\bibinfo {volume} {107}},\
  \bibinfo {pages} {156803} (\bibinfo {year} {2011})}\BibitemShut {NoStop}%
\bibitem [{\citenamefont {Park}\ \emph {et~al.}(2013)\citenamefont {Park},
  \citenamefont {Kim}, \citenamefont {Lee},\ and\ \citenamefont
  {Han}}]{Park2013}%
  \BibitemOpen
  \bibfield  {author} {\bibinfo {author} {\bibfnamefont {J.-H.}\ \bibnamefont
  {Park}}, \bibinfo {author} {\bibfnamefont {C.~H.}\ \bibnamefont {Kim}},
  \bibinfo {author} {\bibfnamefont {H.-W.}\ \bibnamefont {Lee}},\ and\ \bibinfo
  {author} {\bibfnamefont {J.~H.}\ \bibnamefont {Han}},\ }\href
  {https://doi.org/10.1103/PhysRevB.87.041301} {\bibfield  {journal} {\bibinfo
  {journal} {Phys. Rev. B}\ }\textbf {\bibinfo {volume} {87}},\ \bibinfo
  {pages} {041301(R)} (\bibinfo {year} {2013})}\BibitemShut {NoStop}%
\bibitem [{\citenamefont {Go}\ \emph {et~al.}(2017)\citenamefont {Go},
  \citenamefont {Hanke}, \citenamefont {Buhl}, \citenamefont {Freimuth},
  \citenamefont {Bihlmayer}, \citenamefont {Lee}, \citenamefont {Mokrousov},\
  and\ \citenamefont {Bl{\"u}gel}}]{Go2017}%
  \BibitemOpen
  \bibfield  {author} {\bibinfo {author} {\bibfnamefont {D.}~\bibnamefont
  {Go}}, \bibinfo {author} {\bibfnamefont {J.-P.}\ \bibnamefont {Hanke}},
  \bibinfo {author} {\bibfnamefont {P.~M.}\ \bibnamefont {Buhl}}, \bibinfo
  {author} {\bibfnamefont {F.}~\bibnamefont {Freimuth}}, \bibinfo {author}
  {\bibfnamefont {G.}~\bibnamefont {Bihlmayer}}, \bibinfo {author}
  {\bibfnamefont {H.-W.}\ \bibnamefont {Lee}}, \bibinfo {author} {\bibfnamefont
  {Y.}~\bibnamefont {Mokrousov}},\ and\ \bibinfo {author} {\bibfnamefont
  {S.}~\bibnamefont {Bl{\"u}gel}},\ }\href {https://doi.org/10.1038/srep46742}
  {\bibfield  {journal} {\bibinfo  {journal} {Scientific Reports}\ }\textbf
  {\bibinfo {volume} {7}},\ \bibinfo {pages} {46742} (\bibinfo {year}
  {2017})}\BibitemShut {NoStop}%
\bibitem [{\citenamefont {Sunko}\ \emph {et~al.}(2017)\citenamefont {Sunko},
  \citenamefont {Rosner}, \citenamefont {Kushwaha}, \citenamefont {Khim},
  \citenamefont {Mazzola}, \citenamefont {Bawden}, \citenamefont {Clark},
  \citenamefont {Riley}, \citenamefont {Kasinathan}, \citenamefont {Haverkort},
  \citenamefont {Kim}, \citenamefont {Hoesch}, \citenamefont {Fujii},
  \citenamefont {Vobornik}, \citenamefont {Mackenzie},\ and\ \citenamefont
  {King}}]{Sunko2017}%
  \BibitemOpen
  \bibfield  {author} {\bibinfo {author} {\bibfnamefont {V.}~\bibnamefont
  {Sunko}}, \bibinfo {author} {\bibfnamefont {H.}~\bibnamefont {Rosner}},
  \bibinfo {author} {\bibfnamefont {P.}~\bibnamefont {Kushwaha}}, \bibinfo
  {author} {\bibfnamefont {S.}~\bibnamefont {Khim}}, \bibinfo {author}
  {\bibfnamefont {F.}~\bibnamefont {Mazzola}}, \bibinfo {author} {\bibfnamefont
  {L.}~\bibnamefont {Bawden}}, \bibinfo {author} {\bibfnamefont {O.~J.}\
  \bibnamefont {Clark}}, \bibinfo {author} {\bibfnamefont {J.~M.}\ \bibnamefont
  {Riley}}, \bibinfo {author} {\bibfnamefont {D.}~\bibnamefont {Kasinathan}},
  \bibinfo {author} {\bibfnamefont {M.~W.}\ \bibnamefont {Haverkort}}, \bibinfo
  {author} {\bibfnamefont {T.~K.}\ \bibnamefont {Kim}}, \bibinfo {author}
  {\bibfnamefont {M.}~\bibnamefont {Hoesch}}, \bibinfo {author} {\bibfnamefont
  {J.}~\bibnamefont {Fujii}}, \bibinfo {author} {\bibfnamefont
  {I.}~\bibnamefont {Vobornik}}, \bibinfo {author} {\bibfnamefont {A.~P.}\
  \bibnamefont {Mackenzie}},\ and\ \bibinfo {author} {\bibfnamefont {P.~D.~C.}\
  \bibnamefont {King}},\ }\href {https://doi.org/10.1038/nature23898}
  {\bibfield  {journal} {\bibinfo  {journal} {Nature}\ }\textbf {\bibinfo
  {volume} {549}},\ \bibinfo {pages} {492} (\bibinfo {year}
  {2017})}\BibitemShut {NoStop}%
\bibitem [{\citenamefont {Go}\ \emph {et~al.}(2018)\citenamefont {Go},
  \citenamefont {Jo}, \citenamefont {Kim},\ and\ \citenamefont {Lee}}]{Go2018}%
  \BibitemOpen
  \bibfield  {author} {\bibinfo {author} {\bibfnamefont {D.}~\bibnamefont
  {Go}}, \bibinfo {author} {\bibfnamefont {D.}~\bibnamefont {Jo}}, \bibinfo
  {author} {\bibfnamefont {C.}~\bibnamefont {Kim}},\ and\ \bibinfo {author}
  {\bibfnamefont {H.-W.}\ \bibnamefont {Lee}},\ }\href
  {https://doi.org/10.1103/PhysRevLett.121.086602} {\bibfield  {journal}
  {\bibinfo  {journal} {Phys. Rev. Lett.}\ }\textbf {\bibinfo {volume} {121}},\
  \bibinfo {pages} {086602} (\bibinfo {year} {2018})}\BibitemShut {NoStop}%
\bibitem [{\citenamefont {Go}\ \emph {et~al.}(2021)\citenamefont {Go},
  \citenamefont {Jo}, \citenamefont {Gao}, \citenamefont {Ando}, \citenamefont
  {Bl\"ugel}, \citenamefont {Lee},\ and\ \citenamefont {Mokrousov}}]{Go2021}%
  \BibitemOpen
  \bibfield  {author} {\bibinfo {author} {\bibfnamefont {D.}~\bibnamefont
  {Go}}, \bibinfo {author} {\bibfnamefont {D.}~\bibnamefont {Jo}}, \bibinfo
  {author} {\bibfnamefont {T.}~\bibnamefont {Gao}}, \bibinfo {author}
  {\bibfnamefont {K.}~\bibnamefont {Ando}}, \bibinfo {author} {\bibfnamefont
  {S.}~\bibnamefont {Bl\"ugel}}, \bibinfo {author} {\bibfnamefont {H.-W.}\
  \bibnamefont {Lee}},\ and\ \bibinfo {author} {\bibfnamefont {Y.}~\bibnamefont
  {Mokrousov}},\ }\href {https://doi.org/10.1103/PhysRevB.103.L121113}
  {\bibfield  {journal} {\bibinfo  {journal} {Phys. Rev. B}\ }\textbf {\bibinfo
  {volume} {103}},\ \bibinfo {pages} {L121113} (\bibinfo {year}
  {2021})}\BibitemShut {NoStop}%
\bibitem [{\citenamefont {Shick}\ \emph {et~al.}(2017)\citenamefont {Shick},
  \citenamefont {Lichtenstein}, \citenamefont {Shapiro},\ and\ \citenamefont
  {J.Kolorenc}}]{ShickHoPt(111)}%
  \BibitemOpen
  \bibfield  {author} {\bibinfo {author} {\bibfnamefont {A.}~\bibnamefont
  {Shick}}, \bibinfo {author} {\bibfnamefont {A.}~\bibnamefont {Lichtenstein}},
  \bibinfo {author} {\bibfnamefont {D.~S.}\ \bibnamefont {Shapiro}},\ and\
  \bibinfo {author} {\bibnamefont {J.Kolorenc}},\ }\href
  {https://doi.org/10.1038/s41598-017-02809-7} {\bibfield  {journal} {\bibinfo
  {journal} {Scientific Reports}\ }\textbf {\bibinfo {volume} {7}},\ \bibinfo
  {pages} {3} (\bibinfo {year} {2017})}\BibitemShut {NoStop}%
\bibitem [{\citenamefont {Locht}\ \emph {et~al.}(2018)\citenamefont {Locht},
  \citenamefont {Kvashnin}, \citenamefont {Rodrigues}, \citenamefont {Pereiro},
  \citenamefont {Bergman}, \citenamefont {Bergqvist}, \citenamefont
  {Lichtenstein}, \citenamefont {Katsnelson}, \citenamefont {Delin},
  \citenamefont {Klautau}, \citenamefont {Johansson}, \citenamefont
  {Di~Marco},\ and\ \citenamefont {Eriksson}}]{Locht2018}%
  \BibitemOpen
  \bibfield  {author} {\bibinfo {author} {\bibfnamefont {I.~L.}\ \bibnamefont
  {Locht}}, \bibinfo {author} {\bibfnamefont {Y.~O.}\ \bibnamefont {Kvashnin}},
  \bibinfo {author} {\bibfnamefont {D.~C.}\ \bibnamefont {Rodrigues}}, \bibinfo
  {author} {\bibfnamefont {M.}~\bibnamefont {Pereiro}}, \bibinfo {author}
  {\bibfnamefont {A.}~\bibnamefont {Bergman}}, \bibinfo {author} {\bibfnamefont
  {L.}~\bibnamefont {Bergqvist}}, \bibinfo {author} {\bibfnamefont {A.~I.}\
  \bibnamefont {Lichtenstein}}, \bibinfo {author} {\bibfnamefont {M.~I.}\
  \bibnamefont {Katsnelson}}, \bibinfo {author} {\bibfnamefont
  {A.}~\bibnamefont {Delin}}, \bibinfo {author} {\bibfnamefont {A.~B.}\
  \bibnamefont {Klautau}}, \bibinfo {author} {\bibfnamefont {B.}~\bibnamefont
  {Johansson}}, \bibinfo {author} {\bibfnamefont {I.}~\bibnamefont
  {Di~Marco}},\ and\ \bibinfo {author} {\bibfnamefont {O.}~\bibnamefont
  {Eriksson}},\ }\href {https://doi.org/10.1103/physrevb.94.085137} {\bibfield
  {journal} {\bibinfo  {journal} {Physical Review B}\ }\textbf {\bibinfo
  {volume} {94}},\ \bibinfo {pages} {1} (\bibinfo {year} {2018})}\BibitemShut
  {NoStop}%
\end{thebibliography}%

\end{document}